\documentclass[12pt,preprint,xdvi,brief_bib]{aastex}

\usepackage{natbib}
\usepackage{graphicx}

\newcommand{\be}{\begin{equation}}
\newcommand{\ee}{\end{equation}}
\newcommand\eq{Eq.}
\newcommand\eqs{Eqs.}
\newcommand\fig{Fig.}
\newcommand\figs{Figs.}

\newcommand{\half}{\hbox{${1\over2}$}}
\newcommand{\rvec}{{\bf r}}

\newcommand{\yhat}{{\bf \hat{y}}}
\newcommand{\zhat}{{\bf \hat{z}}}
\newcommand{\bhat}{{\bf \hat{b}}}
\newcommand{\xvec}{{\bf x}}
\newcommand{\vvec}{{\bf v}}
\newcommand{\bvec}{{\bf B}}

\newcommand\DEFT{{\tt DEFT}}
\newcommand\EBTEL{{\tt EBTEL}}

\begin{document}

\title{A Model for the Origin of High Density in Loop-top X-ray Sources}

\author{D.W. Longcope and S.E. Guidoni}
\affil{Department of Physics, Montana State University, \\
  Bozeman, MT 59717}


\begin{abstract}
Super-hot looptop sources, detected in some large solar flares, are compact sources of HXR emission with spectra matching thermal electron populations exceeding 30 megakelvins.  High observed emission measure, as well as inference of electron thermalization within the small source region, both provide evidence of high densities at the looptop; typically more than an order of magnitude above ambient.   Where some investigators have suggested such density enhancement results from a rapid enhancement in the magnetic field strength, we propose an alternative model, based on Petschek reconnection, whereby looptop plasma is heated and compressed by slow magnetosonic shocks generated self-consistently through flux retraction following reconnection.  Under steady conditions such shocks can enhance density by no more than a factor of four.  These steady shock relations (Rankine-Hugoniot relations) turn out to be inapplicable to Petschek's model owing to transient effects of thermal conduction.  The actual density enhancement can in fact exceed a factor of ten over the entire reconnection outflow.  An ensemble of flux tubes retracting following reconnection at an ensemble of distinct sites will have a collective emission measure proportional to the rate of flux tube production.  This rate, distinct from the local reconnection rate within a single tube, can be measured separately through flare ribbon motion.  Typical flux transfer rates and loop parameters yield emission measures comparable to those observed in super-hot sources.
\end{abstract}

\date{Draft: \today}

\section{Introduction}

Solar flares are best known for releasing large amounts of magnetic energy.  They are equally remarkable, however, for large increases in plasma density accompanying the energy release.  During a large flare the coronal density can increase by two or three orders of magnitude, from an ambient value like $n_e\sim10^9\,{\rm cm}^{-3}$, to $n_e\sim10^{11}$ or $10^{12}\,{\rm cm}^{-3}$ \citep[see eg.][]{Moore1980b}.  The temperature of this dense plasma is typically one order of magnitude larger than ambient conditions ($T_{\rm fl}\sim20$ MK) so the magnetic field must confine plasma pressure three to four orders of magnitude above normal.  The coronal magnetic field is strong enough to confine this high-pressure plasma in the two directions perpendicular to field lines, but the plasma is unconfined along the field lines.  It is commonly believed that transport of mass and energy along the field lines drives a process known as chromospheric evaporation to supply the extra mass responsible for the density enhancement \citep{Antiochos1978}.  Energy is added to the chromosphere causing it to expand into the corona, thereby diluting the coronal energy over more mass, lowering its temperature but raising its density.

In addition to hot loops, filled by chromospheric evaporation, some flares include an over-dense ``knot'' of plasma at the loop apex known as {\em looptop source}
\citep[LTS, see review by][]{Fletcher1999}.  It is possible that many more flares have LTSs that cannot be imaged by current methods limited in dynamic range \citep{Petrosian2002}.
An LTS is generally assigned a high plasma density due to its radiative output in hard X-rays (HXRs) by which the sources are most commonly imaged.  Its location atop the loop\footnote{Some investigators prefer the term ``above-the-looptop'' in recognition of its separation from loops visible in soft X-ray or EUV images.  In all likelihood the LTS is {\em at the top of} some magnetic loop, even if that loop is not (yet) visible at longer wavelengths.}  suggests the LTS is a direct manifestation of magnetic energy release \citep{Masuda1994}.  If they are thus assumed to arise early in the energy release sequence, it would seem problematic to invoke evaporation as the source of excess density in LTSs.  Moreover, the source appears physically separated from the chromosphere \citep{Jiang2006}, suggesting its mass is not supplied from there.

A subset of LTSs appear to be dense enough that their HXR emission arises from an electron population which is convincingly {\em thermal} (i.e.\ Maxwellian).  The temperatures of these {\em super-hot} (SH) sources typically exceed $30$ MK and sometimes reach $50$ MK \citep{Lin1981,Lin1985,Kosugi1994,Nitta1997}.  One indication of their density comes from the emission measure (EM) of the thermal brehmsstrahlung, typically exceeding 
$10^{48}\,{\rm cm}^{-3}$.  Attributing a large filling factor (often unity) to the apparent source provides density values as high as $n_e\sim 10^{11}$ \citep[a smaller filling factor would demand an even higher density]{Veronig2006,Longcope2010,Caspi2010}.  Independent evidence for these high densities comes from the inference of a Maxwellian electron distribution, supported at high confidence by high resolution HXR spectra \citep{Lin1981,Longcope2010,Caspi2010}.  
Electrons must collide with one another numerous times to fully thermalize.  Accommodating this within a source extending only $10$ Mm requires electrons with mean free paths $\ell_{e}\la1$ Mm along the field line.  Electrons at a typical SH temperature of $T_e=35$ MK  have a collisional mean free path, 
\be
  \ell_{e} ~\simeq~ 100\,{\rm Mm}\left({n_e\over 10^9\,{\rm cm}^{-3}}\right)^{-1} ~~,
\ee
along the magnetic field line, so thermalization within the LTS demands $n_e\ga10^{11}\,{\rm cm}^{-3}$.

Generally, LTSs show a {\em non-thermal} electron population ({\em e.g.}\ a power-law), similar to footpoint sources, during a flare's impulsive phase \citep{Alexander1997}.  Superhot thermal sources typically appear {\em after} the impulsive phase and persist $\sim10$ minutes or more, far longer than the conductive cooling time \citep{Jiang2006}.   Some HXR spectral analyses show the SH source co-existing with a second thermal population at $T_e\sim 20$ MK and much higher emission measure \citep[5--20 times higher,][]{Longcope2010,Caspi2010}.  This lower-temperature component closely matches the soft X-ray light-curve, from GOES for example, commonly attributed to chromospheric evaporation.  This re-affirms a role for the SH component as a flare stage {\em prior} to evaporation, thus demanding a different mechanism for enhancing its density.

To date, little effort has been devoted theoretically explaining the high densities in LTSs.  Non-thermal sources are sometimes assumed to be filled by chromospheric evaporation driven by impact of precipitating particles \citep{Veronig2004}.  Doing so tacitly assumes an energy release process able to persist long enough {\em on a single field line} to interact with its own effects: chromospheric evaporation.  This seems at odds with the favored energy release mechanism, magnetic reconnection, whereby a single field line will reconnect once in  a single instant.  Resolution of this puzzle seems to require a self-consistently coupled model of particle acceleration and magnetic reconnection; such a model is still still being pursued.

Since SH looptop sources are thermal, and show little evidence for non-thermal particles, they might be understood without an understanding of particle acceleration.  Most widely accepted models of magnetic reconnection, such as that of \citet{Petschek1964}, use fluid equations and are therefore directly applicable to SH sources.  Several LTS models proposed so far have invoked an increase in magnetic fields strength, a ``collapsing trap'', to enhance the density and temperature \citep{Somov1997,Karlicky2004,Caspi2010}.   To match observed densities these often require field strengths to increase by factors of 30--100.   These models were not based on reconnection solutions, in fact, assuming a field strength {\em increase} would seem to contradict spontaneous reconnection scenarios where energy is released by {\em reducing} magnetic energy.  The notable exception is the fast magnetosonic termination shock which sometimes forms where the reconnection outflow jet encounters an ``obstacle'', such as the arcade of previously reconnected loops.

Fast magnetosonic termination shocks have been a common feature in reconnection-driven flare models \citep{Forbes1983,Forbes1986,Tsuneta1997,Aurass2002,Vrsnak2005}.  The magnetic field strength increases across a shock of this type, making it suitable for Fermi particle acceleration \citep{Tsuneta1998}.  The density also increases across the shock, but typically by no more than a factor of two \citep{Forbes1986,Vrsnak2005}.  This modest density enhancement (far short of the factor of 100 often seen in loop-top sources) coupled with the low emission measure expected in a small structure, makes the fast magnetosonic termination shock an unlikely candidate for a collapsing trap or a LTS.

The principal element in Petschek's fast reconnection model is a 
{\em slow magnetosonic shock}, or slow shock, which heats and accelerates the 
plasma.\footnote{Petschek's 1964 paper contains several pioneering elements which were  extensively studied in subsequent investigations under the generic term ``Petschek reconnection''.  
It derived a steady external solution matched to a resistive internal solution to obtain its well-known reconnection rate.   This solution was improved and generalized over following decades 
\citep{Sonnerup1970,Vasyliunas1975,Soward1982,Priest1986}.  The solution is, 
however, modified when the electric field is not resistive or is unsteady 
\citep{Heyn1996,Nitta2001}.  Petschek's work was also the first to recognize the generation of slow shocks by reconnection, which turn out to be inevitable in most fast reconnection modes
\citep{Semenov1983,Erkaev2000}.  It is to this aspect of Petschek's work which we exclusively refer hereafter.}
Since field strength {\em decreases} across shocks of this type they are unsuitable for Fermi acceleration.  They do, however, compress the plasma considerably more than the fast magnetosoic termination shock can.  Reconnection between {\em anti-parallel} fields generates slow shocks at the switch-off limit, wherein the maximum possible density enhancement is 2.5 \citep{Forbes1989}.  Reconnecting skewed fields (i.e.\ any angle other than anti-parallel) can, however, result in density enhancements as great as four, according to standard shock models.  This compressed and heated plasma forms a long jet extending away from the reconnection site which may have considerably greater emission measure than the small fast magnetosoic termination shock.  For these reasons it seems plausible for LTSs to be manifestations of the plasma heated and compressed by slow shocks according to Petschek's reconnection model.

Recently \citet{Longcope2010} modeled a SH-LTS observed by RHESSI using slow shocks from a 3-dimensional time-dependent reconnection theory.  The post-shock temperature follows directly from the angle between reconnecting fields lines, which follows in turn from a model of the pre-flare magnetic field.  \citet{Longcope2010} found the observed SH temperature $T\simeq40$ MK to be consistent with the angle from their pre-flare magnetic model: $\Delta\theta=70^{\circ}$ (anti-parallel fields have $\Delta\theta=180^{\circ}$).  The time-dependent model assumed reconnection occurred sporadically in numerous small {\em patches} within the current sheet separating the skewed fields.  Each patch produced a single flux tube whose subsequent retraction through the sheet generated a plug of SH plasma between traveling slow shocks.  Each plug emitted for $\sim 8$ sec before confining inflows ceased and its own pressure ``disassembled'' it.  The single observed LTS was actually composed of $\sim30$ separate, unresolved plugs in flux tubes created by separate patches across the sheet.  The plug collection had a total emission measure proportional to the rate of patch-production which was in turn proportional to the mean reconnection rate.  The latter was measured using the motion of chromospheric flare ribbons and found to be consistent with the emission measure of the observed LTS.  Each reconnected flux tube was later observed at lower temperature as a distinct post-flare loop in TRACE 171\AA\ images.  The rate of loop appearances was also deemed consistent with the the inferred patch-production rate.

While the compression ratio is greater for a slow shock than for the fast magnetosoic termination shock (4 {\em vs.} 2), it is still significantly below the level inferred for observations of looptop sources: 10--100.  This fact forced \citet{Longcope2010} to assume a very high ambient density 
($n_{e0}\simeq 8\times10^{10}\,{\rm cm}^{-3}$) in order to match the observed emission measure.  This same problem faces all models invoking shocks to compress plasma.  The aforementioned  compression ratios, used by all analytical reconnection models, come from standard MHD shock models, based on conservation laws across a steady-state jump \citep[see][for example]{Priest2000}.  One possible exception occurs when radiation can cool the plasma faster than compression heats it --- so called {\em radiative shocks} \citep{Xu1992}.  Unfortunately, the radiative loss function is so low at SH temperatures that even a density of $n_e=10^{12}\,{\rm cm}^{-3}$ cannot produce effective cooling 
($\tau_{\rm rad}\simeq8$ minutes!).  Thus it would seem impossible to invoke a radiative shock in a SH looptop source.

Shocks generated by magnetic reconnection are, however, so far from steady state conditions that standard compression ratios, so often used, are not actually applicable.  Thermal conduction along the magnetic field line will generate a {\em heat front} ahead of the shock \citep{Thomas1944,Grad1951,Germain1960,Forbes1986b,Kennel1988,Yokoyama1997}.  Since thermal conduction does not affect net energy conservation, the heat front does not change the steady-state density ratio.
Under coronal conditions thermal conductivity is extremely large and the heat front would be enormous by the time it achieved steady state --  far larger than any conceivable flare loop \citep{Guidoni2010}.  This forces the conclusion that solar flare shocks never achieve their steady state.  Recent simulations of these shocks by \citet{Guidoni2010} show density enhancements by factors 6--8 during the transient phase following reconnection.  More careful study, herein presented, reveals there is no upper bound to the transient density enhancements and that factors of 10--100 are reasonable for a large solar flare.  We therefore propose that the large plasma densities in LTSs are the direct result of compression in slow shocks of the fast reconnection liberating the energy.

One advantage of the model here proposed is that it directly couples the density enhancement to the energy release.  In models of fast magnetic reconnection, magnetic energy is liberated primarily by shortening field lines following topological change within a very small diffusion region.  Except in cases of purely anti-parallel reconnection, the field strength decreases only slightly and thus the release cannot be interpreted as 
{\em field annihilation}.  The field line shortens at the local Alfv\'en speed which is far larger than the hydrodynamic sound speed ($\beta\ll1$).  The plasma within the shortening flux tube is therefore compressed at {\em super-sonic} speeds, generating the slow shocks which heat the plasma.  This is a time-dependent process on any given field line, so the shock behavior is transient rather than steady.  Compression and heating, i.e.\ the slow shock, is thus a primary effect of releasing magnetic energy, while the fast magnetosoic termination shock is a secondary effect caused by abruptly {\em halting} the energy release.  We propose here that LTSs are manifestations of the primary effect: slow shocks.

We present our model by first reviewing (in \S 2) how magnetic energy is converted to heat when fast reconnection occurs at a current sheet.  We show that similar predictions are made by two-dimensional steady-state models, such as Petschek's, as by transient patchy reconnection models.  The latter can be modeled as a simple shock tube problem.  In \S 3 we use the shock-tube problem to demonstrate how thermal conduction leads to density enhancement far beyond what Rankine-Hugoniot relations predict.  We find a simple relation between the shock-tube mach number and the maximum density enhancement.  We next use thin flux tube models to investigate how much density enhancement could occur within a reconnecting current sheet.  In \S 5 we apply these single-flux tube results to a flare consisting of numerous flux transfer episodes.  We show how the $EM$ of the SH-LTS scales with the rate of flux transfer.   The constant of proportionality is determined by the angle between field lines reconnected at the current sheet.  Section 6 then considers how the energy released by reconnection drives chromospheric evaporation to generate a second, cooler component of the plasma.  Thus  while the chromosphere contributes most the the $EM$ in a flare, it is not responsible for the LTS.

\section{Density increase from reconnection shocks}

\subsection{The equilibrium current sheet}

\begin{figure}[htbp]
\plotone{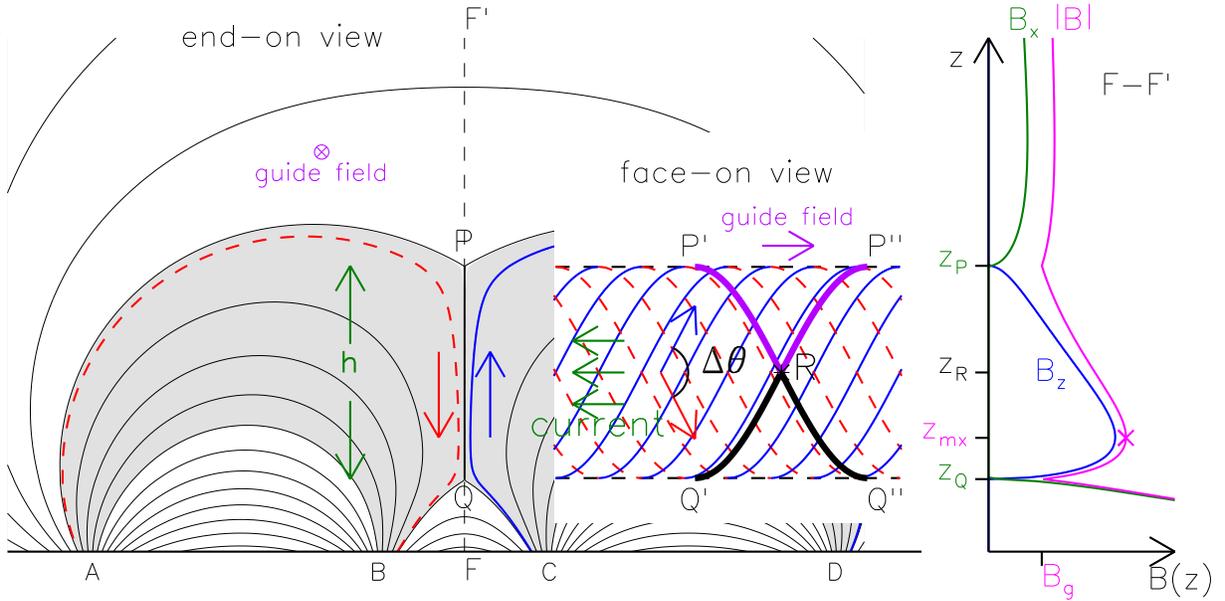}
\caption{A current sheet whose reconnection releases energy.  A quadrupolar field is anchored to sources, $A$, $B$, $C$ and $D$.  In the {\sf end-on} view (left), an equilibrium current sheet extending between {\sf Y}-points at $P$ and $Q$, separates downward field lines (red dashed) connecting $A$ to $B$ from upward (blue solid) connecting $C$ to $D$.  Viewed from {\sf face-on} (inset) the current sheet separates field lines differing by angle $\Delta\theta$.  The strength of the vertical ($B_z$, blue) and horizontal ($B_x$, green) field  components along the vertical $F$--$F'$ are plotted on the right.  The magnitude ($|\bvec|$, magenta) includes the guide field ($B_y$) component.}
	\label{fig:rx_demo}
\end{figure}

Fast magnetic reconnection in models such as that of \citet{Petschek1964}, or subsequent investigators \citep{Forbes1983,Yokoyama1997,Birn2001}, occurs at a discontinuity in the magnetic field called a current sheet (CS).  Figure \ref{fig:rx_demo} shows a simple example of such a CS in a two-dimensional quadrupolar magnetic field \citep{Priest1975}
anchored to four photospheric source regions labeled $A$--$D$.  The coronal field contains free energy  because its field lines interconnect the sources differently than would the lowest-energy field, a potential field.  Here $A$ and $B$ are connected by flux in excess of that required by the potential field; the excess flux, $\Delta\Psi$, is shaded grey in the figure.  The excess magnetic energy can be decreased by a reconnection electric field within the CS breaking an $A$--$B$ (red dashed) and $C$--$D$ (blue solid) field line to create new field lines connecting $A$ to $D$ and $C$ to $B$.  This will decrease the flux excess $\Delta\Psi$ thereby decreasing the free energy.

When the two-dimensional equilibrium includes a field component in the ignorable direction, a {\em guide field} ($B_y$, magenta), the current sheet separates field lines which are not exactly anti-parallel but make a finite angle $\Delta\theta$ (see ``{\sf face-on}'' inset).  Reconnection at point $R$ creates a new field line $AP'RP''D$, the $A$--$D$ line, and  $CQ'RQ''B$ as the $C$--$B$ line.  These are not the shortest possible lines between their end-points.  For example, the section $Q'RQ''$ (thick black) can be shortened to a straight segment $Q'Q''$, along the lower edge of the CS, by pulling the bend through the CS.  In general such reconfiguration would be hindered by other field lines within the equilibrium.    A current sheet, however, represents a ``crack'' in the magnetic field through which field lines may pass unimpeded.  This unimpeded shortening is the basis of all fast magnetic reconnection scenarios, but was posed in an equivalent form first by \citet{Petschek1964}.

All other things being equal, shortening field lines will reduce the net magnetic energy given by the integral
\be
  W _M~=~ {1\over8\pi}\int\limits_{z>0} |\bvec(\xvec)|^2d^3x 
  ~=~ {1\over8\pi}\int d\Phi\int|\bvec|\,d\ell
  ~=~ \int d\Phi\,{\delta W_M\over\delta\Phi}~~,
  	\label{eq:W}
\ee
where the outer integral in the final two expressions is over all field lines composing the corona, and the inner integral, $\delta W_M/\delta \Phi$, is over the individual field lines.  To reduce the total magnetic energy one must decrease the contributions of all flux tubes on average.  This is done by either decreasing their lengths or their average field strengths, or both.  Changing a field line from $Q''RQ'$ to the straighter $Q''Q'$ will clearly do the former.  Since the field strength is minimum along the CS edge ($Q$) moving the field line there will also decrease the average field strength.  Current sheet reconnection therefore liberates magnetic energy in both ways.

Prior to any reconnection the current sheet is part of a stable magnetostatic equilibrium (stable to ideal perturbations, which do not change magnetic field line topology).  This means that its magnetic energy is a minimum and thus its field lines are, in a collective sense, as short as possible: $AP'Q''B$ represents the shortest possible path between $A$ and $B$ for that particular field line.  Shortening that field line would require deforming and lengthening so many neighboring field lines that the net energy would increase.  It is clear that the central segments, $P'Q''$ and $Q'P''$, are relatively straight paths between the end-points when compared to the bends, $P'RP''$ and $Q'RQ''$, resulting from reconnection.  

That shortening a magnetic field line can {\em also} decrease its field strength is not obvious and warrants further comment.  The coronal magnetic field strength tends 
to {\em decrease} with height, so one expects retracting downward would {\em increase} rather than decrease field strength.  This is in fact the case for the {\em arcade} field below the CS ($z<z_Q$ in \fig\ \ref{fig:rx_demo}) and probably describes the later phase of a flare.     It is therefore unclear whether such loops were themselves releasing magnetic energy (their length decrease overwhelming the field increase) or were consuming energy being released elsewhere.

The situation is very different within the CS, $z_Q<z<z_P$.  The tips of the current sheet, $Q$ and $P$, are Y-type null points where the field components perpendicular the the current (i.e.\ the reconnection components shown in green and blue on the right of \fig\ \ref{fig:rx_demo}) vanish.  The magnetic field strength is thus a minimum at each tip: $|\bvec|=B_g$.  Within the region, $z_Q<z<z_{\rm mx}$, the field strength increases with height making it possible to both retract and weaken the field simultaneously.  In addition to its more favorable field-strength profile, the region within the CS offers the most favorable fractional decrease in length: from $Q'RQ''$ to $Q'Q''$; there appears to be less potential for shortening the legs of the post-flare loops, segments $CQ'$ and $Q''B$.  These factors combine to make the CS the region where most of the magnetic energy conversion will occur.  All viable reconnection models are set within the CS and this is the region to which we will confine further consideration.

Shortening a flux tube might also decrease its volume thereby providing a means of raising its average density.  The volume of a flux tube, per unit flux, is
\be
  {\delta V\over \delta\Phi} ~=~ \int{d\ell\over|\bvec|} ~~.
\ee
Decreases in length and field strength contribute in opposite senses and may therefore lead to either an increase or decrease in total volume as a result of energy release.  Moreover, a flux tube anchored in the chromosphere is coupled to a reservoir from which it may change its own mass independent of the volume change.  This is how evaporation can increase the density in the flux tube over long time scales.

The main causes for density increase, at least initially, are slow magnetosonic shocks which are dynamical consequences of very rapid field line retraction.  Dynamical evolution of shortening flux tubes has been treated using several seemingly different models.  In the earliest of these, the steady-state, two-dimensional model of \citet{Petschek1964}, the reconnection electric field occurs along the infinite line $z=z_R$ within a current sheet without guide field 
($\Delta\theta=180^{\circ}$).  The model was generalized by \citet{Soward1982b} to include a guide field and thereby admit $\Delta\theta<180^{\circ}$.  In a further generalization studied by \citet{Linton2006} and \citet{Longcope2009}, the electric field is localized in the erstwhile ignorable direction ($y$).  A transient instance of such ``patchy'' reconnection produces two isolated flux tube, such as the black and violet curves, $Q'RQ''$ and $P'RP''$ respectively, shown in \fig\ \ref{fig:rx_demo}.  These were found to evolve approximately as thin flux tubes, subject to their own magnetic tension.  The layers of flux separated by the CS confine the flux tubes but otherwise let them evolve undisturbed.  The density evolution turns out to be very similar in the two-dimensional, steady state model and the three-dimensional, transient model.

\subsection{Thin flux tube retraction}

We illustrate this generic post-reconnection dynamics with solutions of thin flux tube equations within a CS.  To simplify the geometry we assume a symmetric current sheet of the Green-Syrovatskii type \citep{Green1965,Syrovatskii1971} between $z_Q=0$ and $z_P=h$.  The vertical field component (also called the reconnection component) varies across the sheet as
\be
  B_{z}(z) ~=~ \pm {2B_{z0}\over h}\,\sqrt{z(h-z)} ~~,
  	\label{eq:Bz_GS}
\ee
where the upper and lower sign correspond to the front ($x>0$) and back ($x<0$) of the sheet respectively.  There is a uniform guide field, $B_{y}=B_g$.  We solve only for that portion of the field line within the CS (the $Q'Q''$ segment) and only until effects of reconnection have reached these ends.  We thereby restrict ourselves to times earlier than effects from the boundary, such as chromospheric evaporation, could play a role.

The dynamical evolution is treated using the thin flux tube dynamics of \citet{Linton2006}, including parallel dynamics added by \citet{Longcope2009} and field-aligned transport added by \citet{Guidoni2010}.  The axis of the tube is a space curve, 
\[
  \rvec(\ell,t)~=~y(\ell,t)\yhat ~+~z(\ell,t)\zhat ~~, 
\]
parameterized by length $\ell$.  The unit tangent vector $\bhat=\partial\rvec/\partial\ell$, is parallel to the internal magnetic field.  The low-$\beta$ tube moves through the current sheet governed by its own dynamics.  The velocity of an element changes due to internal gas pressure gradients, magnetic curvature, external magnetic pressure gradients and viscosity according to \citep{Guidoni2011}, but gravitational forces are neglected,
\be
  \rho{d\vvec\over dt} ~=~ -\bhat{\partial p\over\partial \ell} ~+~
  {B^2\over4\pi}{\partial \bhat\over\partial\ell}  ~-~ {1\over4\pi}\nabla_{\perp}B^2 ~+~
  B{\partial\over\partial\ell}\left( {\mu\over B}\,\bhat\bhat\cdot{\partial \vvec\over\partial\ell}
  \right) ~~.
  	\label{eq:TFT_mom}
\ee
Here $\nabla_{\perp}$ is the component of the gradient perpendicular to $\bhat$.
While evolution of the curve changes the field's direction, its strength is set by pressure balance across the tube.  The external pressure is provided by the flux outside the current sheet making it a fixed function of position
\be
  B(y,z) ~=~ \sqrt{B_y^2+B_z^2} ~=~ \sqrt{B_{g}^2 ~+~ 4B^2_{z0}\,{z\over h}
  \left(1-{z\over h}\right)} ~~~~,~~~~ 0<z<h ~~.
\ee
The tube is assumed to be small enough not to affect the sheet's equilibrium by its motion.   The dynamic viscosity $\mu$ depends strongly on temperature $T=(\bar{m}/k_{\rm B})p/\rho$, where $\bar{m}$ is the mean mass per particle 
and $k_{\rm B}$ is Boltzmann's constant.  Its temperature dependence is given by the collisional Spitzer-Harm form
\be
  \mu ~\simeq~ 0.12~{\rm dyne\cdot sec}\, (T/10^6\, {\rm K})^{5/2} ~~.
  	\label{eq:mu}
\ee
This is small enough that it is often ignored, but it is essential for resolving the shocks which develop within the flux tube.

We solve these equations using a Lagrangian numerical code \citep[\DEFT,][]{Guidoni2010} which follows the tube's mass elements and finds the mass density $\rho$ using the elongation between pairs of elements and the cross-sectional area $\delta A=\delta\Phi/B$.  The resulting density undergoes an evolution satisfying \citep{Guidoni2011}
\be
  {d\ln\rho\over dt} ~=~{d \ln B\over dt} ~-~ {\partial (\bhat\cdot\vvec)
  \over\partial\ell}~+~ \vvec\cdot{\partial \bhat\over\partial\ell} ~~.
  	\label{eq:TFT_cont}
\ee
The first term on the right hand side (rhs) reflects the density enhancement from lateral compression as the tube moves into stronger magnetic field: $\rho\propto B$; it will move into stronger field if the reconnection site is above the sheet's midpoint ($z_{\rm R}>h/2$).  The second term reflects axial compression either from shortening or from shock compression.  The third produces compression when radius of curvature is decreased through perpendicular motion.

The pressure of a given mass element evolves according to an energy equation
\be
  {dp\over dt} ~=~ {5\over 3}p\,{d\ln\rho\over dt} ~+~{2\over3}\,\mu\left(
  \bhat\cdot{\partial \vvec\over\partial\ell}\right)^2 ~+~ {2\over3}B\,{\partial\over\partial\ell}
  \left({\kappa\over B}{\partial T\over\partial\ell}\right) ~~.
  	\label{eq:TFT_erg}
\ee
The first term on the rhs reflects adiabatic work, with $\gamma=5/3$, the second is viscous heating and the third is thermal conduction.\footnote{Since heating occurs relatively slowly through viscosity the thermal conduction is generally small and we find it unnecessary to use any form of conductive flux limiter \citep[see][]{Longcope2010b}.}   The thermal conductivity, $\kappa$, has temperature dependence identical to the viscosity and can be expressed as $\kappa = (k_{\rm B}/\bar{m})\mu/{\rm Pr}$, where ${\rm Pr}=0.012$ is the Prandtl number using Spitzer collisional transport.\footnote{The Prantdl number is very small because viscous stress is transmitted through random ion motion, while most thermal energy is carried by the lighter, faster electrons (in this single-fluid treatment, a.k.a.\ MHD, both species have the same temperature)}  Since we consider only a flux tube, all transport processes are automatically field aligned; we consider no viscous or thermal transport across the magnetic field.  Radiative losses are neglected here because they are negligible at the high temperatures and short times scales of interest; we return to the point below.

\begin{figure}[htbp]
\plotone{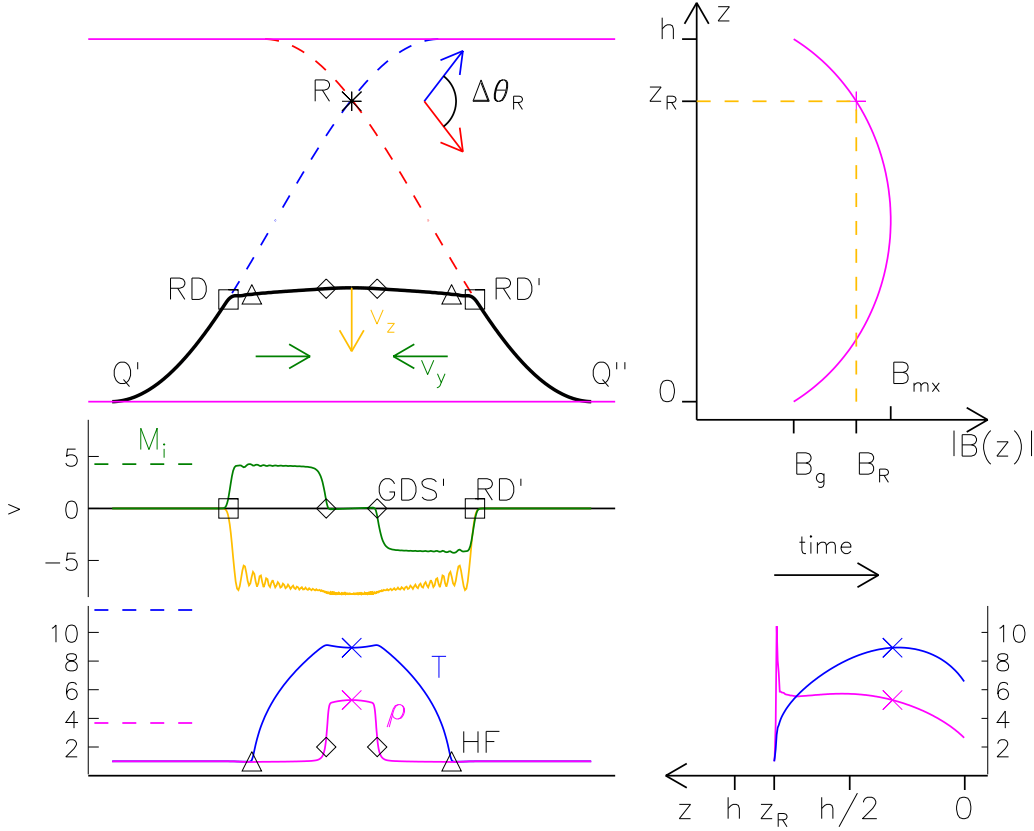}
\caption{A thin flux tube simulation of post-reconnection relaxation.  The initial field lines are shown in the upper left panel as red and blue dashed curves.  These reconnect at point $R$ (asterisk) and relax to the solid black curve at a later time.  Fluid within the tube moves downward ($v_z$ orange) and inward ($v_y$ green).  The profiles of these velocity components are shown as curves below, normalized to $c_{{\rm s}0}$, and the peak 
of $v_y$, corresponding to $M_i$, is represented as a green dashed horizontal line.  Both velocity components are initiated by magnetic tension at the rotational discontinuity, {\sf RD'} (square).  The horizontal velocity is reduced at the gas dynamic shock, {\sf GDS'} (diamond).  The density ($\rho$, magenta) and temperature ($T$, blue) are plotted below this, normalized to their ambient values.  Horizontal dashed lines of the same color show the Rankine-Hugoniot values corresponding to the $M_i$ from above.  The density changes at the slow shock; the temperature increases within the heat fronts ({\sf HF}, triangles).  The central value ($\times$s) undergo a time-evolution plotted in the lower right panel against height, $z$ (increasing leftward).}
	\label{fig:deft}
\end{figure}

The flux tube evolution begins {\em after} the reconnection electric field has created a bent flux tube, $Q'RQ''$ in \fig\ \ref{fig:deft}.  (The other section, $P'RP''$ evolves according to identical dynamics, so we disregard it.)  The equations are initialized with uniform density and pressure and a configuration, $\rvec(\ell,0)$, tracing curve $Q'RQ''$.  The reconnection electric field can in principle dissipate energy, heat ions or electrons, or even accelerate particles.  All these possible effects are, however, ignored when the tube is initialized with uniform pressure and density.   We choose to neglect these effects on the premise that the dissipation occurs within such a small volume that it cannot have a significant energetic effect on the subsequent dynamics.  This same approximation, neglecting direct dissipation such as Ohmic heating, was used by Petschek and subsequent investigators of fast magnetic reconnection.

Figure \ref{fig:deft} shows the numerical solution of \eqs\ (\ref{eq:TFT_mom}),
(\ref{eq:TFT_cont}), and (\ref{eq:TFT_erg}), by \DEFT.  The initial bend at point $R$ decomposes into two rotational discontinuities, $RD$ and $RD'$, and two gas-dynamic shocks, $GDS$ and $GDS'$, as described in \citet{Longcope2009}.  The rotational discontinuities propagate along the erstwhile equilibrium field lines at the local Alfv\'en speed.  They change the direction of the magnetic field, but not its strength or any plasma properties.  Tension at the rotational discontinuity accelerates plasma along the bisector moving the plasma both downward ($v_z$ in orange, this is the retraction) and {\em inward} ($v_y$ in green).  The section between them is slightly curved owing to the variation in this speed within the CS \citep[see][]{Guidoni2011}, but is basically horizontal.  Thus the inward flow is actually {\em parallel} to the axis of the flux tube within the $RD$--$RD'$ segment.  While the Lorentz force (i.e.\ tension) accelerates only perpendicular to the local field, subsequent changes to the field direction renders the flow parallel.

The inward parallel flow generated by the rotational discontinuities resembles a hydrodynamic shock tube.  The field bends to approximately horizontal through an angle $\Delta\theta/2$.  The field strength $|\bvec|$ is the same as in the equilibrium field and the plasma properties, including initial plasma beta, $\beta_0$, are those of the pre-reconnection ambient state.  The parallel flow thus has a hydrodynamic Mach number\footnote{This should be confused with the {\em Alfv\'en} Mach number often used in the reconnection literature to characterize the reconnection rate.  $M_i$ is normalized to the sound speed $c_s$ and characterizes the speed of field line retraction {\em following} reconnection.}
 \citep{Longcope2009,Longcope2010}
\be
  M_i ~=~ \sqrt{24\over5\beta_0}\sin^2(\Delta\theta/4) ~~.
  	\label{eq:Mi}
\ee
 The solution in \fig\ \ref{fig:deft} begins with reconnection at $\Delta\theta_R=106^{\circ}$ and $\beta_{0R}=0.02$, which would generate an inflow at $M_i=3.1$.  In the middle of the CS, $z=h/2$, where the field is stronger 
($\beta_0=0.014$) and the angle is greater ($\Delta\theta=120^{\circ}$) the flow could reach $M_i=4.6$.  By the later time shown in the figure the actual parallel flow has a Mach number $M_i=4.2$ (see middle panel).

These inflows should not be confused with the slow ``inflows'' feeding the reconnection in steady state reconnection models.  In the more conventional {\sf end-on view}, these appear flowing toward the current  sheet, where they are accelerated, by rotational discontinuities, to form the ``outflow jet'', directed away from the reconnection site (upward and downward).  The inflows shown in \fig\ \ref{fig:deft} are actually part of the same ``outflow jet'', having been accelerated by the rotational discontinuities.  Since they are part of the outflow jet, they are also present  in steady-state models with a guide field, but the component directed along the ignorable direction (i.e.\ the inflow, $v_y$) is not often discussed.  In the less-conventional {\sf face-on view}, shown in the figure it is evident that these flows are natural consequences of field line shortening.

\subsection{Shocks}

The supersonic, inward, parallel flows from each rotational discontinuity collide in the center creating gas dynamics shocks where the parallel flow speed drops to zero (green curve in middle panel of \fig\ \ref{fig:deft}).  Both the density and temperature increase across such a shock, resulting in a hot dense central plug (magenta and blue curves in the bottom panel).  

\begin{figure}[htb]
\epsscale{0.75}
\plotone{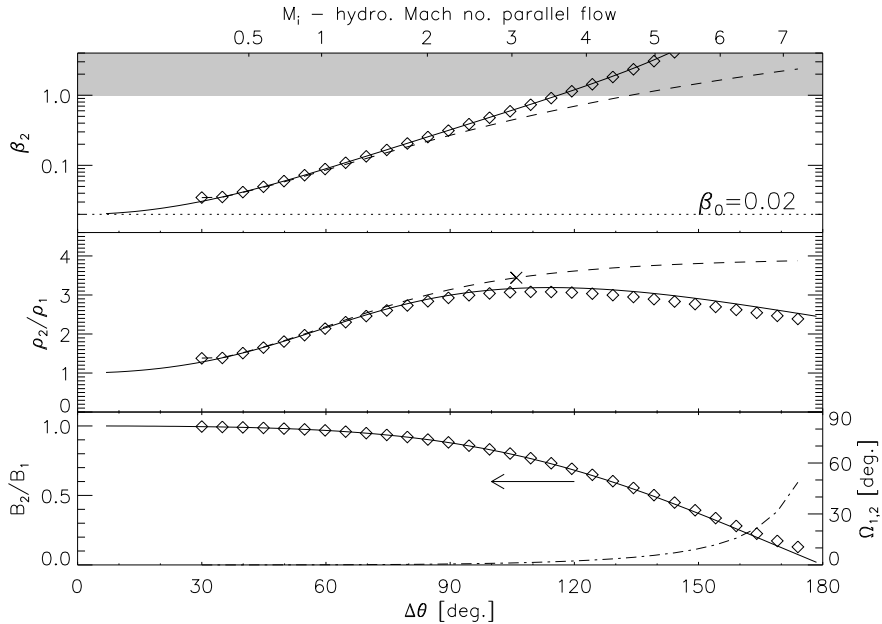}
\caption{Jump conditions across steady shocks for different reconnection angles 
$\Delta\theta$, according to different models of post-reconnection dynamics.  The top panel shows the plasma $\beta$ of the post-shock material, and the middle shows the density ratio.  Solid curves are from the 2d steady state model of \citet{Soward1982b}, the diamonds are for the 1d Riemann problem of \citet{Lin1994} and the dashed for the thin flux tube model of \citet{Longcope2009}.  The latter corresponds to a shock tube for inflow Mach number given by \eq\ (\ref{eq:Mi}), plotted along the top axis.  A $\times$ marks the conditions at point $R$ of the run shown in \fig\ \ref{fig:deft}.  The bottom panel shows the jump in magnetic field strength (against the left axis) and the change in field direction (against the right axis) for the two MHD models.  In the thin flux tube these are fixed to be $B_2/B_1=1$ and $\Omega_{1,2}=0$.}
	\label{fig:cmp}
\end{figure}

The gas dynamic shocks differ in several significant respects from the slow magnetosonic shocks of the classical Petschek model, especially its anti-parallel version \citep[$\Delta\theta=180^{\circ}$,][]{Longcope2009}.  When the two-dimensional steady model includes a guide field, however, there are two sets of shocks: rotational discontinuities enclosing an outflow jet, inside of which are slow magnetosonic shocks where heating occurs 
\citep{Soward1982b,Forbes1989,Skender2003}.  For cases of non-trivial guide field ($\Delta\theta<150^{\circ}$) and low $\beta$, the slow shocks are very nearly parallel shocks resembling gas dynamic shocks.  Figure \ref{fig:cmp} shows only a modest decrease in magnetic field strength ($B_2/B_1\simeq1$) and very small change in field direction, $\Omega_{1,2}$, across 
the slow shock.   These cases look far less like Petschek's original model with a hot outflow jet, than like the thin flux tube model shown in \fig\ \ref{fig:deft}.  The greatest difference is a superficial one: a steady outflow jet (Petschek) is replaced by a single plug moving downward and expanding outward.  The latter becomes more akin to the former when multiple reconnection events are combined to generate a sequence of plugs.

Many applications of reconnection models, either Petschek's or the patchy thin flux tube model, predict the density and temperature enhancement from shock jump conditions, sometimes called Rankine-Hugoniot relations.  Such relations follow from conservation laws and from the assumption that the shock is steady when viewed in a co-moving frame \citep{Courant1948}.  Under these conditions the density, temperature and magnetic field strength jump from an upstream value (subscript 1) to a downstream value (subscript 2).  The ratio of downstream to upstream values depends on the Mach number $M_i$ which follows in turn from the angle $\Delta\theta$.  This angle is a parameter of the pre-reconnection current sheet in both thin flux tube models and steady, two-dimensional models; it should not be confused with the opening angle of the outflow jets when the latter solutions are viewed {\sf edge-on}.  

Figure \ref{fig:cmp} shows the density ratio predicted by Rankine-Hugoniot relations for two-dimensional steady state reconnection (solid) and thin flux tube dynamics (dashed).  A third version of the relations, obtained from a one-dimensional Riemann problem beginning with bent field lines \citep{Lin1994}, is plotted using diamonds.  This simplified problem captures the non-resistive evolution following reconnection and has been shown to reproduce properties of two-dimensional steady state reconnection \citep{Lin1999b}.

The Rankine-Hugoniot density ratio for the shock tube (dashed curve in the middle pane of \fig\ \ref{fig:cmp}), 
\be
  {\rho_{_{\rm RH}}\over\rho_0} ~=~ {\rho_2\over\rho_1} ~=~
  {8M_i^2+4M_i\sqrt{4M_i^2+9}+9\over
  2M_i^2 + M_i\sqrt{4M_i^2+9} + 9} ~~,
  	\label{eq:rho_RH}
\ee
has a well-known asymptote $\rho_2/\rho_1\to4$ at large $M_i$ 
(i.e.\ $\Delta\theta\to180^{\circ}$ and $\beta_0\to0$).  In this same limit the slow shocks of the MHD models (2d steady and 1d Reimann problem) approach switch-off shocks --- the limit originally used by Petschek.  A switch-off shock annihilates the magnetic field ($B_2/B_1\to0$) and has a limiting density ratio $\rho_2/\rho_1\to2.5$ \citep{Forbes1989}.  For less severe angles, say $\Delta\theta<120^{\circ}$, the slow shock barely deflects the magnetic field, ($\Omega_{1,2}<3^{\circ}$) and therefore resembles a gas dynamics shock.  Since this is the same as in the shock tube all three models show very similar behavior in this regime.  There is still some decrease in field strength across the slow shock which is naturally absent from the shock tube.    The former thus has a slightly lower density enhancement than the latter, although both are above the switch-off limit of 2.5.  The very large post-shock $\beta$ (top panel) is dominated by plasma pressure increase since the magnetic pressure decrease is small (it is zero for the shock tube).

The density enhancement in the time-dependent solution does not, however, match the steady Rankine-Hugoniot value of\ \eq\ (\ref{eq:rho_RH}).  The lower panel of \fig\ \ref{fig:deft} contains dashed horizontal lines showing the jumps predicted by Rankine-Hugoniot conditions for $M_i=4.2$ (the observed Mach number).  The central density notably exceeds this value; in fact it exceeds the maximum possible ratio of 4.  This is due to a ``transient'' density overshoot, and temperature undershoot, which has been observed in previous time-dependent thin flux tube simulations \citep{Guidoni2010,Guidoni2011}.   During the earliest phase of the retraction (see lower right panel of \fig\ \ref{fig:deft}) there is a brief period where the density ratio reaches 10: far above the theoretical maximum of Rankine-Hugoniot shocks.

This density overshoot is a direct consequence of transient behavior resulting from thermal conduction.  When transport is dominated by thermal conduction (i.e.\ when ${\rm Pr}\ll1$) the steady shock profile includes a {\em heat front}  ahead of the primary density jump 
\citep{Thomas1944,Grad1951,Germain1960,Kennel1988,Forbes1986b,%
Xu1992,Yokoyama1997}.  These are evident in the temperature profiles of \fig\ \ref{fig:deft}, where the leading edge is marked by a triangle.  For typical coronal parameters the steady state size of the heat front is so large that is cannot be achieved during retraction \citep{Guidoni2010}.  Thus the entire relaxation process occurs out of steady state as the heat front continues expanding toward its asymptotic size.

It seems that the actual density enhancement caused by reconnection shocks follows from transient shock evolution in the presence of thermal conduction rather than from the steady-state Rankine-Hugoniot values.    The actual enhancement can be several times larger than the steady Rankine-Hugoniot values.  This is a significant departure from standard practice where Rankine-Hugoniot jump conditions have been widely used to estimate emission from fast magnetic reconnection \citep{Petschek1964,Cargill1982,Vrsnak2005,Longcope2010}.  While we have observed the discrepancy in thin flux tube models, these closely match MHD models in other respects so we hope to apply thin flux tube results generally.

\section{Thermal conduction transients in shocks}

The density overshoot can be studied thoroughly in a simplified, one-dimensional shock tube problem.  Equations of continuity, momentum and energy conservation, along the linear tube, parameterized by length $s$, are
\begin{eqnarray}
  {\partial \rho\over\partial t} &=& -{\partial (\rho u)\over \partial s} ~~,\\
    {\partial (\rho u)\over\partial t} &=& -{\partial (\rho u^2)\over \partial s} 
    ~-~{\partial p\over \partial s} ~+~{\partial\over\partial s}\left(\mu 
    {\partial u\over \partial s} \right)~~,\\
    {\partial p\over\partial t} &=& -u{\partial p\over \partial s} ~-~{5\over3}\,p
    {\partial u\over\partial s}
    ~+~ {2\over3}\mu\left({\partial u\over \partial s}\right)^2 
    ~+~ {2\over3}
    {\partial\over\partial s}\left(\kappa {\partial T\over \partial s} \right) ~~,
\end{eqnarray}
using $\gamma=5/3$ as before.
Viscosity $\mu$ and conductivity $\kappa$ are computed using Spitzer-Harm form, \eq\ (\ref{eq:mu}), as they are in the thin flux tube model.  This simple set of equations is equivalent to \eqs\ (\ref{eq:TFT_mom}), (\ref{eq:TFT_cont}) and (\ref{eq:TFT_erg}), for a static, straight flux tube ($\ell\to s$ and $\bhat\cdot\vvec\to u$).

The equations are solved within the interval, $0<s<L$, with a rigid wall  at the left boundary ($s=0$): $u=0$ and $\partial\rho/\partial s=\partial p/\partial s=0$.  The wall represents the central stagnation point in the horizontal segment ($RD$--$RD'$) of the thin flux tube solutions, and values there will henceforth be referred to as ``central''.  A steady, uniform inflow is introduced at the right boundary, $u(L)=-M_i\,c_{{\rm s}0}$, $p(0)=p_0$ and $\rho(0)=\rho_0$, where $c_{{\rm s}0}=\sqrt{\gamma p_0/\rho_0}$ is the ambient sound speed and $M_i$ is the inflow Mach number.  This represents the rotational discontinuity at which the parallel flow is created and its location is irrelevant for the gas dynamic shock; in practice it is repositioned further right whenever an effect from the wall approaches it.

The natural length scale in the problem
\be
  \ell_{i0} ~=~ {\mu_0\over \rho_0 c_{{\rm s}0}} ~=~ 43\,{\rm km}\,
  \left({T_0\over 10^6\, {\rm K}}\right)^2\,\left({n_{e0} \over 
  10^9{\rm cm}^{-3}}\right)^{-1} ~~,
\ee
is related to the ion mean free path.  The corresponding time scale,  related to the ion collision time, is 
\be
  \tau_{i0} ~=~{\ell_{i0}\over c_{{\rm s}0} } ~=~ 0.25\,{\rm sec}\,
  \left({T_0\over 10^6\, {\rm K}}\right)^{3/2}\,\left({n_{e0} \over 
  10^9{\rm cm}^{-3}}\right)^{-1}~~.
\ee

The system is initialized with uniform properties, $\rho(s,0)=\rho_0$ and $p(s,0)=p_0$.  The velocity is initialized with a smooth step,
\be
  u(s,0) ~=~ -M_i\,{\rm tanh}(s/\lambda ) ~~,
  	\label{eq:u0}
\ee
whose gradient scale is denoted $\lambda$.  It is necessary to position the right boundary, $L\gg\lambda$, in order that the initial condition be consistent with the boundary condition.

\begin{figure}[htbp]
\epsscale{1.0}
\plotone{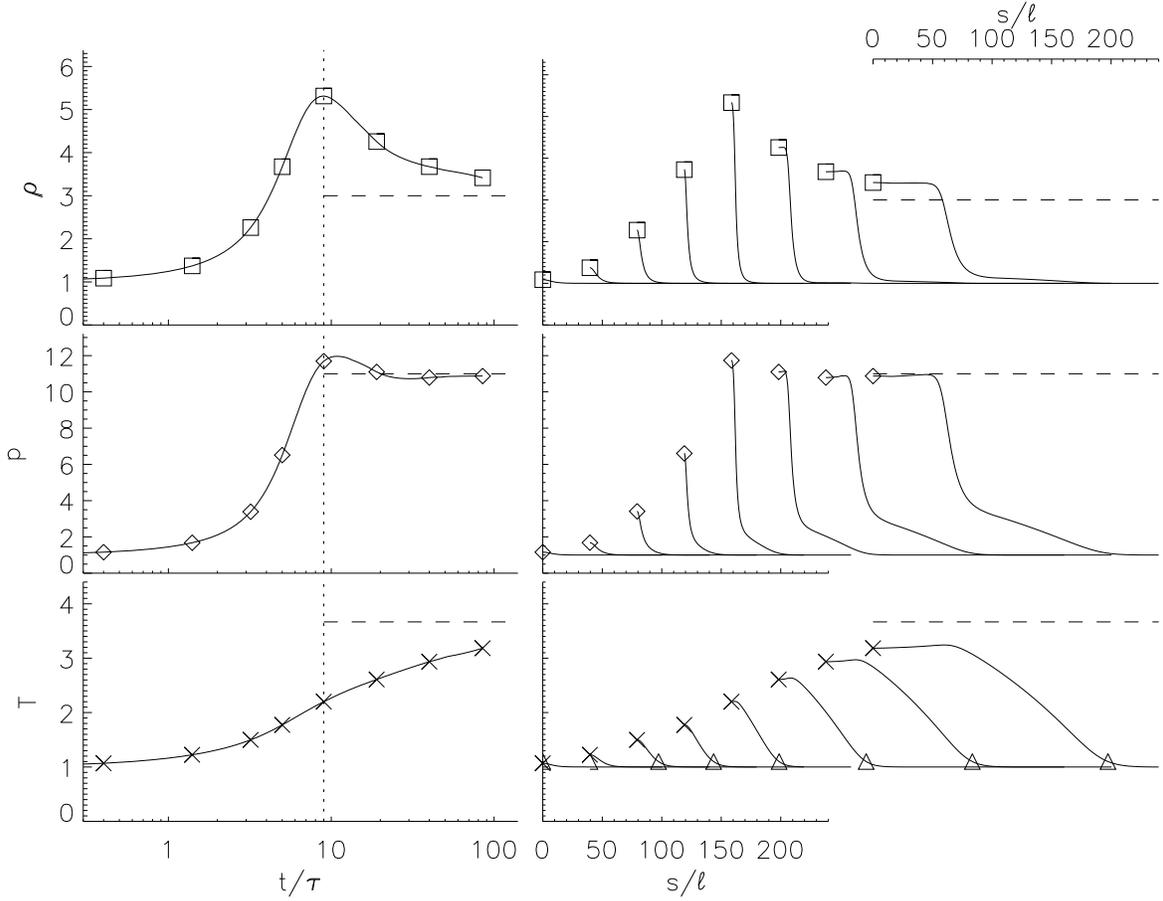}
\caption{The evolution of the shock tube solution for $M_i=2.0$, $\lambda=10\ell_{i0}$.  The right column shows the profiles, $\rho(s,t)$ (top), $p(s,t)$ (middle), and $T(s,t)$ (bottom) at a sequence of times.  Successive profiles are displaced rightward, and symbols show $s=0$ for each profile.  An axis along the bottom of each shows $s/\ell_{i0}$ at $t=0$, while a second along the top of the upper plot shows the same coordinates as they pertain to the final (right-most) time  ($t=100\tau_{i0}$).  $\triangle$s on the bottom plot show the extent of the heat front.  The left column shows the central value, $\rho(0,t)$, $p(0,t)$, and $T(0,t)$, over time (scaled to $\tau_{i0}$).  Symbols correspond the times of the profiles on the right.  The vertical dotted line shows the time of peak density.  Horizontal dashed lines on all panels show the Rankine-Hugoniot values for each quantity.}
	\label{fig:M2}
\end{figure}

The solutions to these equations, such as the example shown in \fig\ \ref{fig:M2}, show features similar to the retracting flux tube of \fig\ \ref{fig:deft}.  Density piles up against the wall ($s=0$) until a pressure is achieved capable of driving a shock into the flow.  This compression raises the temperature of the high-density plug, but thermal conduction leads to extended heat fronts (triangles) in front of the shock.  The pressure achieves its final (Rankine-Hugoniot) value promptly, and thereafter remains fairly constant.  The peak temperature rises more slowly due to thermal diffusion spreading the heat into the broad region behind the front.  The density therefore {\em overshoots} its final value and decreases thereafter.  The fully developed conduction front is extremely large, and is therefore achieved only after a long time.  As it is approached the temperature climbs and density falls -- both rather slowly.

Since the tube length $L$ is irrelevant, $\mu$ defines the length scale $\ell_{i0}$, and the Prantdl number is fixed (${\rm Pr}=0.012$) the only two genuinely free parameters of this system are the Mach number, $M_i$, and the initial shear length $\lambda$.  Together they define the ballistic collapse time, $\tau_{\rm b}=\tau_{i0}\ell_{i0}/\lambda M_i$, taken for the initial velocity shear to steepen into a shock.  

\begin{figure}[htbp]
\epsscale{1.0}
\plotone{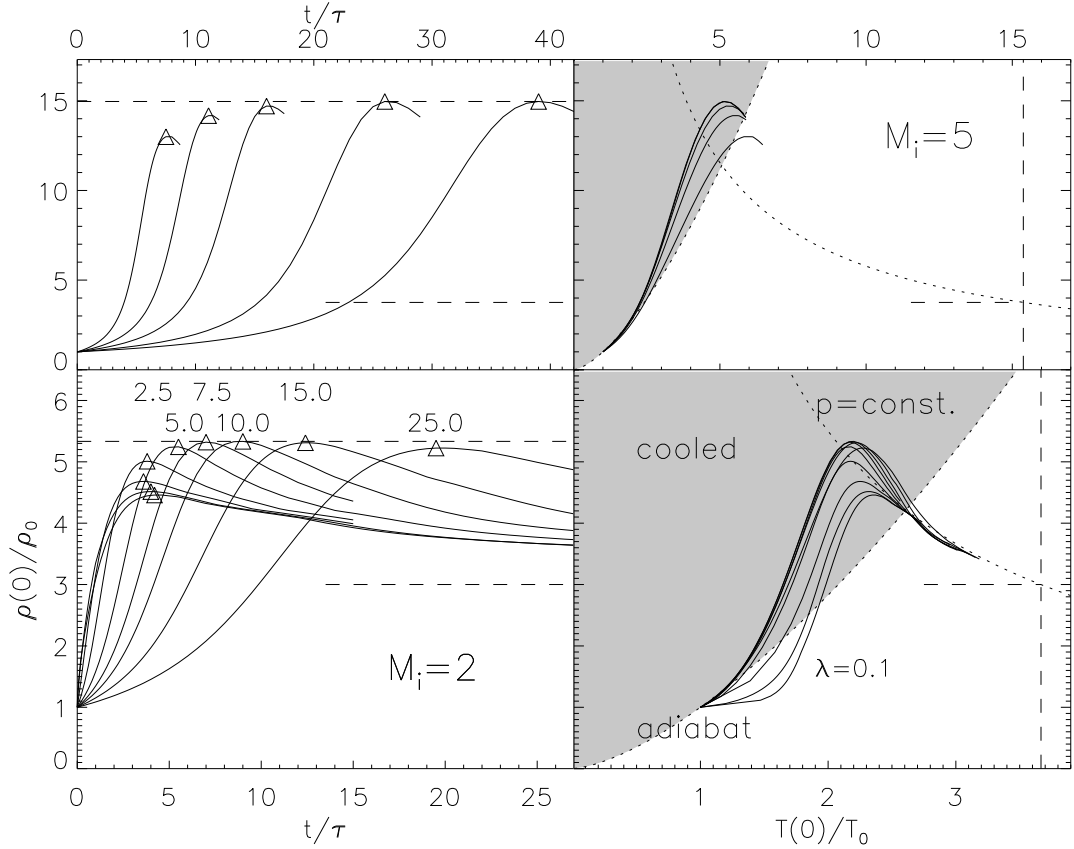}
\caption{Density evolution for cases of $M_i=2$ (bottom) and $M_i=5$ (top).  Left panels show the central density, $\rho(0,t)$, versus time for initial conditions characterized by different values of $\lambda/\ell_{i0}$.  Triangles denote point of peak central density.   In the lower left panel these values are (reading peak values clockwise from the lowest) $\lambda/\ell_{i0}=0.1$, $0.3$, $1$, $2.5$, $5$, $7.5$, $10$, $15$ and $25$;  in the upper left panel they are $\lambda/\ell_{i0}=25$, $40$, $60$, $100$ and $150$.  The right panels show the same evolution plotted as
$\rho(0,t)$ {\em vs.} $T(0,t)$.  Vertical and horizontal dashed curves are the Rankine-Hugoniot values.   The dotted curve dividing grey from white regions is the adiabatic curve through the initial state; the grey region has entropy below initial.  The other dotted curve is the locus with pressure matching the Rankine-Hugoniot state.}
	\label{fig:M25}
\end{figure}

The size of the initial velocity gradient, $\lambda$, relative to $\ell_{i0}$, determines two slightly different patterns in the evolution of the central density (see \fig\ \ref{fig:M25}).  For small initial gradients, $\lambda\gg\ell_{i0}$, the viscous heat generated during the steepening is more than compensated by the thermal conduction.  As a consequence the central entropy decreases slightly during shock formation, causing $\rho(0)$ to evolve sub-adiabatically.  In the opposite limit, $\lambda\ll\ell_{i0}$, the initial velocity shear is rapidly dissipated by viscosity before the density can respond.  This leads to a rapid increase in central entropy which is later offset by thermal conduction.  The pressure from this initial heating hinders the later collapse, so cases with $\lambda\ll\ell_{i0}$, tend to reach lower peak central densities than do the smaller initial gradients.

The density overshoot is a consequence of the initial growth of the heat front by thermal conduction.  Except for cases of steep initial gradients ($\lambda\ll\ell_{i0}$) the initial evolution is nearly adiabatic, and then becomes slightly sub-adiabatic as conduction cools the central fluid, as just described.  The collapse time, $\tau_{\rm b}$, increases with $\lambda$ causing peak density to occur later, as shown in \fig\ \ref{fig:M25}.  The peak value achieved is, however, little different than for modest values of $\lambda$.  For low mach number of $M_i=2$ (bottom row) the largest density enhancement occurs for $\lambda=10\ell_{i0}$, at 
$t=9\simeq 2\tau_{\rm b}$.  For the large Mach number, $M_i=5$ (top row), the peak is at $\lambda=100\ell_{i0}$, at $t=27\simeq 1.5\tau_{\rm b}$.

\begin{figure}[htb]
\epsscale{0.75}
\plotone{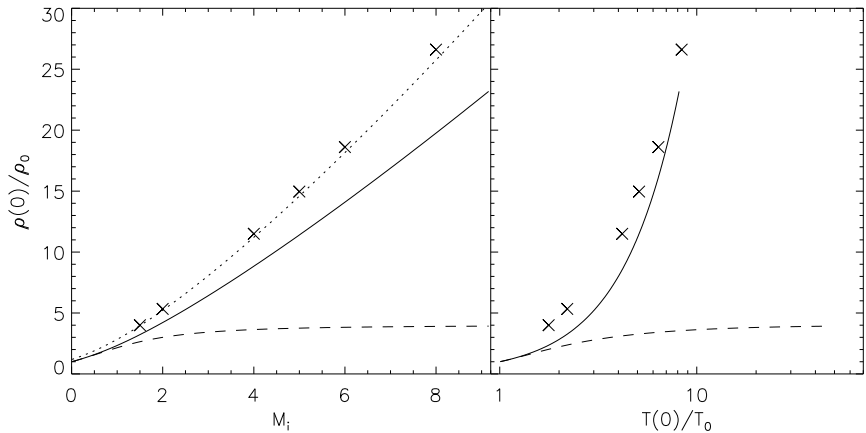}
\caption{Peak density enhancements, ${\rm max}[\rho(0,t)/\rho_0]$, for shock tube solutions with mach numbers $M_i$.  Crosses are solutions with optimal gradient scale $\lambda$.  The solid curve shows the adiabatic enhancement from \eq\ (\ref{eq:rho_ad}).  Dashed curves are the Rankine-Hugoniot enhancements from \eq\ (\ref{eq:rho_RH}).  Left panel shows the enhancement {\em vs.} Mach number; right panel shows it {\em vs.} the temperature at time of peak density.}
	\label{fig:mach_sweep}
\end{figure}

The initial conduction is so effective at cooling the central plasma that the peak density occurs just above the intersection of the adiabatic curve, $p\sim\rho^{5/3}$, and the final (Rankine-Hugoniot) pressure $p_{_{\rm RH}}$.  This point is at the crossing of the two dotted curves on the right panels of \fig\ \ref{fig:M25}.  The adiabatic density enhancement
\be
  {\rho_{\rm ad}\over \rho_0} ~=~ \left({p_{_{\rm RH}}\over p_0}\right)^{3/5} ~~,
  	\label{eq:rho_ad}
\ee
is thus a lower bound to the maximum density overshoot.  Figure \ref{fig:mach_sweep} shows how this enhancement increases rapidly with $M_i$.  For $M_i>1$ this is significantly greater than the steady-state (Rankine-Hugoniot) enhancement (dashed) which asymptotes to 4.  The maximum enhancement for runs with optimal $\lambda$ ($\times$) lie above the adiabatic curve.  An empirical fit $\rho_{\rm mx}=1.33\,\rho_{\rm ad}$ (dotted) fit the data well.

\section{Post-reconnection enhancements}

The thin flux tube model of post-reconnection retraction differs from the shock tube problem in several respects.  First it is initialized at rest, rather than with a prescribed velocity profile like \eq\ (\ref{eq:u0}).  The inflow is generated not by a right hand boundary, but at a rotational discontinuity moving horizontally away from the center at some fraction of
the local Alfv\'en speed.  The rotational discontinuity travels along a field line with varying field strength and direction.   This variation leads, in turn, to variation in $\beta_0$ and $\Delta\theta$, and thus in $M_i$ according to \eq\ (\ref{eq:Mi}).  The gas dynamic shocks will therefore come from inflow with time-dependent properties.   Finally, the variation in field strength causes its own variation in density according to the first term on the rhs of \eq\ (\ref{eq:TFT_cont}); this latter effect was extensively investigated by \citet{Guidoni2011}.

In spite of these differences, the central density enhancement and heat fronts of the thin flux tube simulations are very similar to those of the shock tube.  This is evident by comparing profiles on the right of \fig\ \ref{fig:M2} to the bottom panel of \fig\ \ref{fig:deft}.  Both cases have heat fronts far ahead of the density jump (slow shock).  In both cases the
central density overshoots its Rankine-Hugoniot value and then approaches it from above, as a direct result of the growing heat fronts.

\begin{figure}[htbp]
\epsscale{0.85}
\plotone{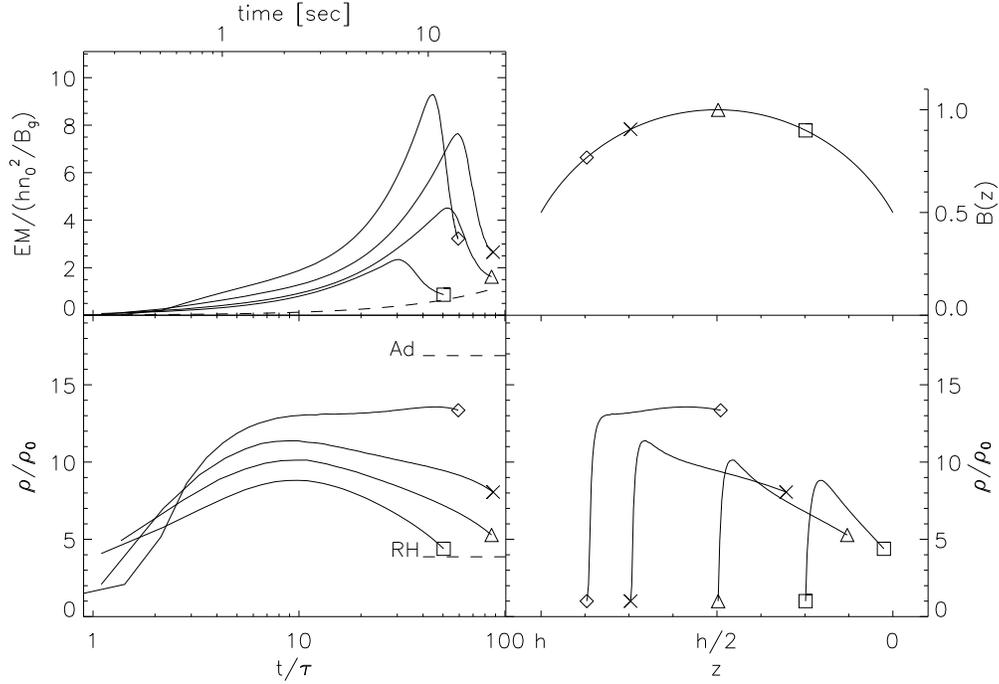}
\caption{Density evolution for thin flux tube solutions in identical current sheets with reconnection initiated at positions, $z_R/h=0.875$ (diamond), $0.75$ ($\times$), $0.5$ (triangle), and $0.25$ (square).  The field strength at the reconnection point is plotted on the upper right curve.  Value of $\beta_R$ was takes to be $\beta_R=6\times10^{-3}$ (diamond), 
$5\times10^{-3}$ ($\times$ and square) and ($3\times10^{-3}$) in order that the initial Mach number was $M_i=7$.  The bottom row shows the evolution of the peak density versus time (left) and height (right; note that $z$ {\em decreases} rightward, which is the sense of the evolution).  The upper left column shows the total emission measure of heated plasma versus time.  The dashed curve shows the result for a steady $M_i=7$ shock.  The bottom time axis is scaled to $\tau_{i0}$; the top axis uses $n_{e0}=3\times10^{9}\,{\rm cm}^{-3}$ and
$T_{0}=2$ MK to convert this to seconds.}
	\label{fig:deft_sweep}
\end{figure}

Figure \ref{fig:deft_sweep} shows several other examples of central density enhancements from thin flux tube solutions.  All occur within identical current sheets: $\Delta\theta_0=120^{\circ}$, and height $h=3\times10^{3}\,\ell_{i0}$. In each case the reconnection is initiated at different positions $z_R$ and the $\beta_R$ was adjusted to keep the initial value of $M_i=7$, according to \eq\ (\ref{eq:Mi}).   In all solutions the central density rises rapidly to a value above the Rankine-Hugoniot limit (the horizontal dashed line on the lower left).  In reconnections closer to the bottom the density falls immediately after its initial rise.  For the highest reconnection point, $z_R=0.875h$ (diamond), the progression into stronger fields and greater angles keeps the density ratio at $\rho/\rho_0\simeq13$ --- far above the Rankine-Hugoniot limit.

When plotted agains $t/\tau_{i0}$ (lower left) the central density shows its further resemblance to the shock tube.  Since most reach their peak before $t=10\tau_{i0}$, even though $M_i=7$, these would seem to resemble shock tube solutions with $\lambda\ll\ell_{i0}$.  This would explain their tendency to reach a peak value below the adiabatic, 
$\rho=\rho_0(p_{_{\rm RH}}/p_0)^{0.6}$, shown as a horizontal dashed curve.  
Other runs reveal that reducing 
$\ell_{i0}$, and thus $\tau_{i0}$, has an effect similar to increasing $\lambda$: peak density is raised still further.  Doing so at high Mach number is numerically challenging due to the very small scale created at the center.

Additional comparison with observations can be made using the emission measure $EM$ of the reconnection-heated plasma.  Thin flux tube solutions yield $EM$ per unit flux in analogy to the field line integral in \eq\ (\ref{eq:W}),
\be
  {\delta EM\over\delta \Phi} ~=~ \int\,n_e^2\,{d\ell\over B} ~~,
\ee
where this integral is only over the heated portion of the tube (i.e. where $T>T_0$).
Time evolution of this quantity is plotted in the upper left panel of \fig\ \ref{fig:deft_sweep}, normalized to $n_{e0}^2h/B_g$.  In each case the $EM$ continues to increase after the density has peaked, owing to the increasing mass swept up by the shock and heat front.  

The $EM$ begins dropping before the simulation ends, as the heat front and rotational discontinuities generate a growing low-density void (rarefaction wave) ahead of the shock \citep{Guidoni2011}.  The dashed line in the upper left of \fig\ \ref{fig:deft_sweep} shows the emission measure predicted by the the simple Rankine-Hugoniot analysis used by \citet{Longcope2010}: density and temperature jumping at a steady jump moving at the Rankine-Hugoniot shock speed.  All cases exceed this prediction due to their much higher density as well as the greater extent of their heat fronts.


\subsection{General behavior}

\begin{figure}[htb]
\epsscale{0.75}
\plotone{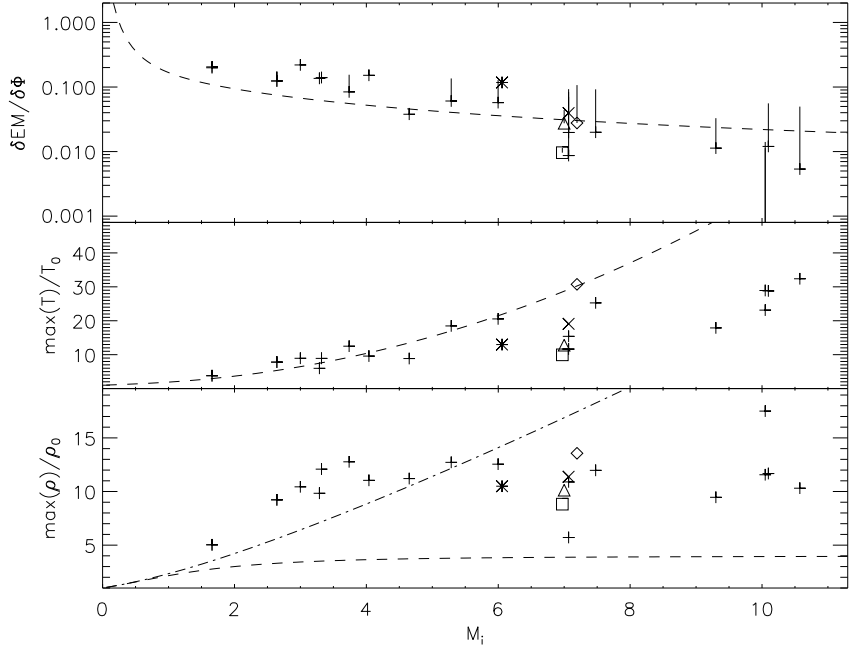}
\caption{Thin flux tube solutions over a range of Mach numbers, $M_i$.  The 4 runs from \fig\ 
\ref{fig:deft_sweep} are designated by the same symbols (diamond, $\times$, triangle and square); the run from the following section is designated by a $*$; all others are plotted by a $+$.  The top, middle and bottom panels plot $\delta EM/\delta\dot{\Phi}$, from \eq\ (\ref{eq:EMphi_dot}), temperature and density enhancement respectively.  The dashed lines correspond to the Rankine-Hugoniot limit from \eqs\ (\ref{eq:EM_RH}) and (\ref{eq:rho_RH}) in the top and bottom panels.  The broken line in the bottom panel is the adiabatic density ratio given by expression (\ref{eq:rho_ad}).  The symbols on the top panel given 
$\delta EM/\delta\dot{\Phi}$, when the simulation ends, often when the heat fronts reach $Q'$ or $Q''$.  Vertical upward lines estimate how high it would be if the simulation could continue until the retraction was complete.}
	\label{fig:smry}
\end{figure}

The general behavior of the thin flux tube model can be gleaned from a composite of many different solutions.  The bottom panel of \fig\ \ref{fig:smry} shows the maximum density enhancment {\em vs.} $M_i$ for 25 different runs.  The collection spans central reconnection angles $90^{\circ}\le\Delta\theta_0\le133^{\circ}$, mean free paths  
$7\times10^{-5}\le\ell_{i0}/h\le7\times10^{-3}$, and reconnection points
$0.75\le z_R\le0.82$ except for the examples shown in \fig\ \ref{fig:deft_sweep} 
(indicated by symbols).  Runs with the highest Mach numbers terminate before complete retraction due to the arrival of the heat fronts at the loop footpoints (this is discussed further below).  This is why plots against $z$ in \fig\ \ref{fig:deft_sweep} (lower right) 
end short of the origin.  A more systematic sample of runs would clearly be useful, and will be attempted in a future investigation.  The present hodge-podge does, however, serve our present purpose by illustrating how typical solutions compare to the Rankine-Hugoniot model, plotted as a dashed line.

The density enhancements in \fig\ \ref{fig:smry} (bottom panel) lie well above the Rankine-Hugoniot value (dashed) in all the myriad runs.  The latter model of density enhancement appears to be needlessly conservative.  The cases with higher Mach numbers tend to have enhancements below the adiabatic limit (broken line, \eq\ [\ref{eq:rho_ad}]).  This is probably due, as remarked above, to the similarity of the thin flux tube solutions to shock tubes solutions with $\lambda\ll\ell_{i0}$.  Density ratios at or above a factor of 10 appear to be the rule, with exceptions most often for low reconnection points (square) or extremely small CS (i.e.\ large $\ell_{i0}/h$).

The maximum observed temperature ratio, plotted in the middle panel of\ \fig\ \ref{fig:smry}, is approximately limited by the Rankine-Hugoniot value (dashed curve).  The runs at highest Mach number require the longest to reach steady shock conditions (i.e.\ Rankine-Hugoniot assumptions), and are the ones stopped before complete retraction.  As a result their temperature maxima tend to fall well below the Rankine-Hugoniot value.  Even in these cases, however, the peak temperature is 10--20 times the starting value, easily eligible for a SH classification.

\subsection{Mean properties: super-position}

The results above apply to a single episode of one post-reconnection flux tube retracting.  We assume a solar flare consists of hundreds of such episodes as flux is transferred through the current sheet at hundreds of separate patches.  The net properties of the flare follow from a super-position of episodes resembling those in \fig\ \ref{fig:deft_sweep}.  If the flux transfer occurs at a mean rate $\dot{\Phi}$, the mean $EM$ of the super-hot source will be
\be
  \langle EM\rangle_{_{SH}}  ~=~ \dot{\Phi}\int {\delta EM\over\delta \Phi}\, dt ~=~
  \dot{\Phi}\times{\delta EM\over\delta\dot{\Phi}} ~~,
  	\label{eq:EMphi_dot}
\ee
where the time integral is over a single episode.
Time integrals of the four curves in \fig\ \ref{fig:deft_sweep} yield $\delta EM/\delta\dot{\Phi}=0.03$, $0.09$, $0.14$ and $0.10$ times the dimensional factor $n_{e0}^2h^2/c_{s0}B_g$, 
for $z_R/h=0.25,\,0.5,\,0.75,\,0.875$.  Even though the highest reconnection point ($z_R=0.875\,h$) reaches the greatest density, its heat fronts reach the footpoints earliest giving it a shorter duration and thus a smaller net emission measure,
 $\delta EM/\delta\dot{\Phi}$, than the reconnection at $z_R=0.75\,h$.  The dimensionalizing factor
\be
  {n^2_{e0}h^2\over c_{s0}B_g} ~=~ 6\times 10^{29}\,{\rm {sec\over cm^3\,Mx}}
  \left({n_{e0}\over 10^9\,{\rm cm}^{-3}}\right)^2\left({h\over 100\,{\rm Mm}}\right)^2
  \left({T_0\over 10^6\,{\rm K}}\right)^{-1/2}\left({B_g\over 10\,{\rm G}}\right)^{-1}~~.
  	\label{eq:EM_fctr}
\ee
gives a reasonable LTS, $\langle EM_{_{SH}}\rangle\sim10^{48}\,{\rm cm}^{-3}$, for 
reconnection at a typical mean rate $\dot{\Phi}=2\times 10^{18}\,{\rm Mx/sec}$ 
\citep{Longcope2010}.   

The top panel of \fig\ \ref{fig:smry} show $\delta EM/\delta\dot{\Phi}$ for all 25 runs in the collection.  In cases where the run terminates before full retraction, an attempt is made to correct the value to a full retraction.  The instantaneous $EM$ at the run's end, which is typically lower than the peak, is extended to the time the tube would have reached $z=0$.  This extrapolation is indicated by a vertical line upward from the value actually observed (symbol).  Since faster heat fronts occur for larger $M_i$ (the right of the plot), those are the cases with vertical extensions.

The net $EM$ can be compared to the prediction of a steady-shock model \citep{Longcope2010}
\be
  {\delta EM_{_{RH}}\over \delta\dot{\Phi}} ~=~ \left({n^2_{e0}h^2\over c_{s0}B_g}\right)\,
  \left\{\,{\rho_{_{RH}}/\rho_0\over 1 - \rho_0/\rho_{_{RH}}}\,{1\over M_i}\,{z_R^2\over h^2}\,
  \cos(\Delta\theta_R/2)\,\tan^2(\Delta\theta_R/4)\,\right\} ~~,
  	\label{eq:EM_RH}
\ee
plotted as a dashed curve on the top panel of 
\fig\ \ref{fig:smry} for $z_R=h/2$ and $\Delta\theta_R=120^{\circ}$.
This would be the $EM$ in a steady 2.5-dimensional model of the 
\citet{Petschek1964} variety.  The inverse scaling with $M_i$ results from the faster retraction speed, relative to the normalizing speed, $c_{s0}$, when $\beta_R$ becomes small (and thus $M_i$ becomes large).  The thin flux tube solutions have somewhat greater $EM$ than the Rankine-Hugoniot model would predict, but appear to share its decrease with increasing $M_i$.

\section{Energetics}

\begin{figure}[htbp]
\epsscale{1.0}
\plotone{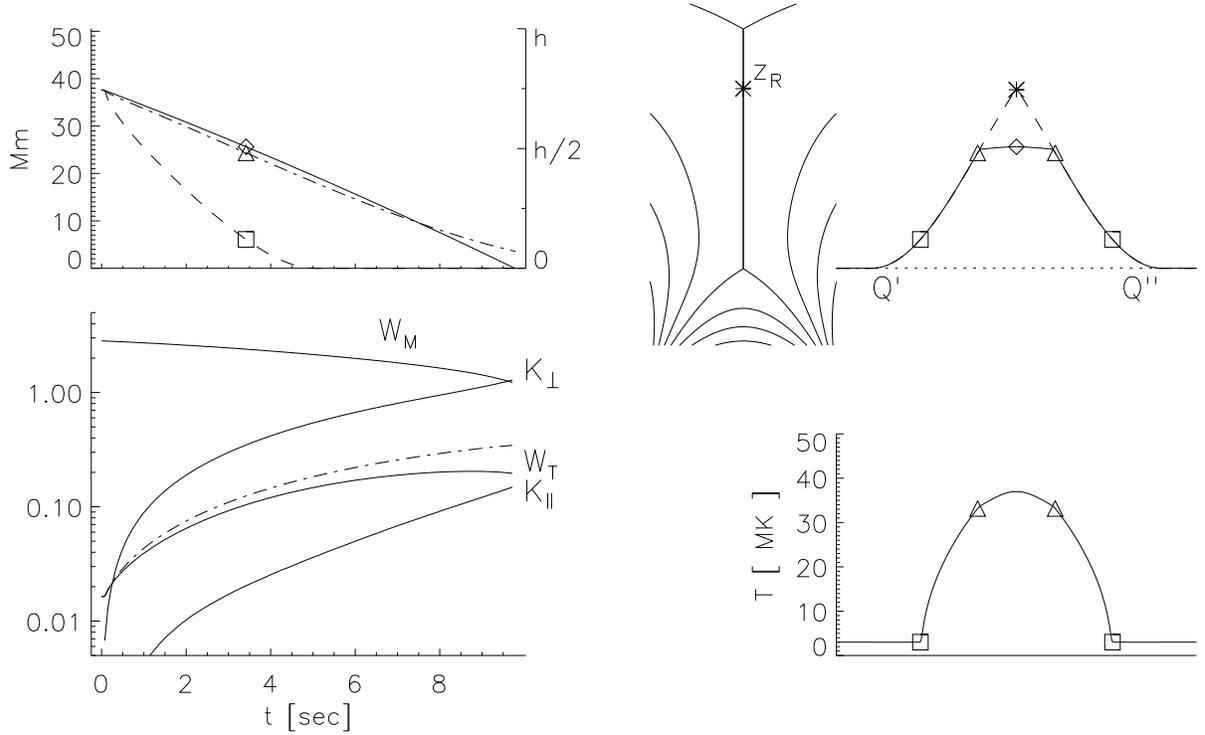}
\caption{A run with $M_i=6$ reconnected at $z_R=0.75\,h$ in a CS with $\Delta\theta_0=120^{\circ}$.  The central panel shows, for illustration purposes, an {\sf end-on} view of the CS, with an asterisk marking the reconnection site.   The {\sf face-on-view} 
of the FT is plotted to the right of this; the dashed curve is the initial state, and solid curve is a later time ($t=3.5$ sec).  Squares and triangles mark the heat fronts and rotational discontinuities respectively, and a diamond marks the tube's center. The temperature profile at this same time is plotted below it on the same scale.  The panel to the left of the {\sf end-on} perspective shows the vertical positions of the heat fronts (dashed with square) rotational discontinuities (broken with triangle) and tube center (solid with diamond) {\em vs.} time (in seconds).  The symbols occur at the time of the solid curve in the upper right.  Below this the various energies are plotted against the same time axis, normalized to the energy minimum, ${\cal E}_0$, given in \eq\ (\ref{eq:E0}).  The solid curves are, $W_M^{\rm (eff)}$, $K_{\perp}$, $W_T$ and $K_{\parallel}$, from top to bottom.  The broken line is $W_T+K_{\parallel}$, which remains confined to the flux tube after retraction has terminated.}
	\label{fig:erg}
\end{figure}

To illustrate the energetics of creating of the SH-LTS, and its subsequent evolution, we consider a specific CS suitable for a compact flare.  The CS is $h=50$ Mm tall with 
$\Delta\theta_0=120^{\circ}$ and a guide field $B_g=86$ G (see \fig\ \ref{fig:erg}).  Reconnection is initiated at $z_R=0.75\,h=37.5$ Mm where $\Delta\theta_R=113^{\circ}$ and $B_R=156$ G.  Properties of this run are plotted with an asterisk on \fig\ \ref{fig:smry}.  The right panels of\ \ref{fig:erg} shows the heat fronts preceding the rotational discontinuities, and the upper left shows that they reach points $Q'$ and $Q''$ at $t=5$ sec., well before the rotational discontinuities do.

The initial tube, $Q'RQ''$ is $L(0)=98$ Mm long, and retraction shortens it to $L_{\rm min}=60$ Mm, the distance between $Q'$ and $Q''$.  Remarkably, while the length is decreased by 40\%, the volume of the flux tube remains virtually unchanged due to weakening of the field.  Thus no density enhancement can be attributed to compression through field line retraction.

The magnetic energy, per unit flux, inside the tube is given by the line 
integral in \eq\ (\ref{eq:W}).  The confining external field does work on the flux tube as it retracts \citep{Longcope2009}, contributing additional energy to an effective potential
\be
  {\delta W_M^{\rm (eff)}\over\delta\Phi} ~=~ {1\over 4\pi}
  \int|\bvec|\,d\ell ~~.
\ee
Since thin flux tube equations, \eqs\ (\ref{eq:TFT_mom})--(\ref{eq:TFT_erg}) are approximately conservative (to $\sim\beta_0$), any decrease in this potential energy appears as either thermal energy or bulk kinetic energy.  In its initial state the flux tube has a potential energy 
$\delta W_M^{\rm (eff)}/\delta\Phi=12\times10^{10}\,{\rm erg/Mx}$.  Shortening and weakening reduces this by 66\% to
\be
  {\cal E}_0 ~=~ 
  {\delta W_{M,0}^{\rm (eff)}\over\delta\Phi} ~=~ {B_gL_{\rm min}\over 4\pi} 
  ~=~ 4\times 10^{10}\,{\rm erg/Mx} ~~.
    	\label{eq:E0}
\ee
Retraction will therefore liberate
\be
  \Delta{\cal E} ~=~ {\delta W_{M}^{\rm (eff)}\over\delta\Phi}\Biggr\vert_{t=0} ~-~ {\cal E}_0
  ~=~ 8\times 10^{10}\,{\rm erg/Mx} ~~
  	\label{eq:DE}
\ee
--- energy available to the power the flare.

The ratio of liberated energy to ${\cal E}_0$ is not equivalent to the more common ratio of free energy to the energy of the potential field.  The equilibrium current sheet carries a net current
\be
  {I_{CS}\over c} ~=~{1\over4\pi}\oint \bvec\cdot d{\bf l} ~=~
  {1\over \pi}{B_{z0}\over h}\int\limits_0^h\sqrt{z(h-z)}\,dz ~=~
  {B_{z0}h\over 8} ~=~ 
  9\times 10^{10}\,{\rm Mx/cm} ~~,
\ee
after using \eq\ (\ref{eq:Bz_GS}) for the vertical field at the CS.  (Translating to SI gives $I_{CS}=9\times 10^{11}$ Amps, typical of pre-flare currents).   Denoting by $\Delta\Psi$ the flux discrepancy giving rise to the current sheet, the flux which must be reconnected (see \fig\ \ref{fig:rx_demo}), the free energy in the equilibrium can be derived from integrating the work required to establish the current sheet: $\Delta W=(I/c)\Delta\Psi/2$.  If the present case has a typical pre-flare flux, $\Delta\Psi=3\times10^{21}$ Mx, its free energy would be $\Delta 
W=3\times10^{32}$ ergs.  The potential field of the entire active region will, however, include more flux, $\Phi_{_{\rm AR}}\sim3\times 10^{22}$ Mx, and thus contain energy about an order of magnitude larger.  Multiplying \eq\ (\ref{eq:DE}) by $\Delta\Psi$ gives twice $\Delta W$ because $\Delta{\cal E}$ quantifies the differential energy release at the full CS 
size $h$.  As $h$ decreases so will $\Delta{\cal E}$; integrating the release over the entire reconnection, $h\to0$, approximately matches $\Delta W$.  

To dimensionalize the plasma response we choose ambient density and temperature $n_{e0}=8\times10^{9}\,{\rm cm}^{-3}$ and $T_0=3\times 10^6$ K so that $\beta_r=0.007$ and $M_i=6.0$ according to \eq\ (\ref{eq:Mi}).  The temperature profile at an early time (see the lower right of \fig\ \ref{fig:erg}) shows the heat fronts (squares) running far ahead of the rotational discontinuities (triangles).  After 5 seconds the fronts have reached the bottom of the CS (points $Q'$ and $Q''$) even though the center of the flux tube is only at the midpoint.  To study evolution beyond this time we include additional straight segments running along the CS edge, for the heat fronts to propagate along \citep{Guidoni2011}.  This permits the solution to be followed until the center of the retracting flux tube (diamond) reaches the bottom of the CS (at $t=10$ sec).  This occurs before the rotational discontinuities reach the bottom, so the tube is not quite straight: $L=1.07\,L_{\rm min}$.  As a result the potential energy is still $22\%$ above 
${\cal E}_0$ (see $W_M$ curve in \fig\ \ref{fig:erg}).

As the tube retracts it converts potential (magnetic) energy into thermal energy and kinetic (per unit flux), given by the field line integrals
\be
  {\delta W_{T}\over\delta \Phi} ~=~ {3\over 2}\int p\,{d\ell \over B} ~~~~~,~~~~~~
  {\delta K\over\delta \Phi} ~=~ {1\over 2}\int \rho |\vvec|^2\,{d\ell \over B} ~~.
\ee
The kinetic energy can be further decomposed into that from motion perpendicular 
($K_{\perp}$) and parallel ($K_{\parallel}$) to the tube's axis $\bhat$.  The lower left panel of \fig\ \ref{fig:erg} shows how all these energies increase as the potential energy decreases.  The majority of the potential energy decrease, $0.78\Delta{\cal E}$, is converted into perpendicular motion: the rapid retraction.  The remaining energy is converted initially into parallel flow 
($K_{\parallel}$) the collision of which generates the gas dynamic shocks which thermalize a fraction.  By the end of this simulation $9\%$ of the liberated energy has been thermalized while $13\%$ remains in parallel flow.  While the thermal energy is concentrated within the central mass concentration, the plasma $\beta$ never exceeds $0.8$.

\section{Density evolution following retraction}

\begin{figure}[htb]
\epsscale{0.75}
\plotone{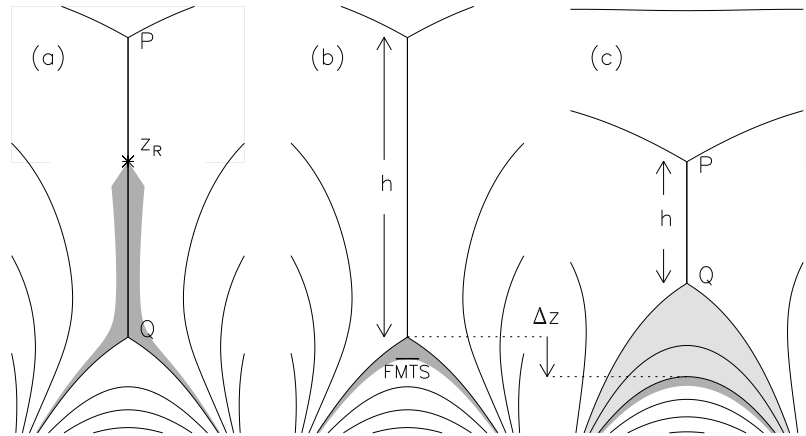}
\caption{Three stages in the post-reconnection relaxation within a current sheet viewed from the {\sf end-on} perspective.  Dark gray shows a particular flux tube at (a) $t=0$ just after reconnection at point $z=z_R$,  (b) $t=t_b$ after being retracted through the current sheet, and (c) $t=t_c$ after additional flux  (light grey) has been reconnected.  The additional reconnection between $t_b$ and $t_c$ decreases the vertical extent, $h$, of the current sheet, moving point $Q$ upward.   The dark grey flux tube moves downward by $\Delta z$ (arrow) in a process known as ``shrinkage''.  Its volume also decreases as its apex field strength increases: ``collapsing trap''.}
	\label{fig:stages}
\end{figure}

The model above concerned only the initial stage in post-reconnection evolution: the reconnected flux tube retracting through the CS.  To place this into the context of the entire flare we briefly discuss the subsequent stages illustrated in \fig\ \ref{fig:stages}.  The downward retraction of the flux tube will be stopped by the closed field lines underlying the current sheet, the arcade (\fig\ \ref{fig:stages}b).  At this point the field line has shortened as much as possible, and its magnetic energy has been converted mostly to bulk kinetic energy in Alfv\'enic downward motion, $K_{\perp}$.  This energy must somehow be dissipated if the flux tube is to join the static arcade.  In a steady-state model this dissipation occurs at a standing fast magnetosonic termination shock shown in the figure \citep{Forbes1983,Forbes1986}.  In a 3d, time-dependent model, such as the thin flux tube, it is more likely to be a traveling shock or a pulse of MHD waves initiated by the impulsive encounter between retracting flux tube and static arcade.  After this occurs the static loop will enter a cooling phase.

The different phases may be intermixed if heat fronts drive chromospheric evaporation during the retraction.  Figure \ref{fig:erg} shows the heat fronts reaching the edge of the CS ($Q'$ and $Q''$) far ahead of the tube itself.  From there they would continue along the arcade legs of the tube eventually reaching the chromospheric footpoints.  Conductive flux into the chromosphere would drive evaporative upflows into the loop.  If the heat fronts were fast enough this upflow could reach the rotational discontinuities and modify the initial stage.  This seems unlikely in the present case where the rotational discontinuities reach the CS bottom only 5 seconds after the heat front.  We therefore neglect this possibility for the moment and return to discuss its implications below.  We prevent any back-reaction from the tube footpoints by adding very long straight sections, as described above.

Assuming the retraction occurs in the absence of chromospheric evaporation allows us to clearly distinguish that evolutionary phase from the subsequent cooling phase, 
\figs\ \ref{fig:stages}b--\ref{fig:stages}c.  In this later phase the loop has achieved magnetic equilibrium within the arcade.  The energy left in the loop from its retraction drives chromospheric evaporation by which radiative cooling is enhanced.
It is this much longer phase that is observed in soft X-ray and EUV images as post-flare loops.  Subsequent retraction episodes will pile additional flux on top at the same time it decreases the CS size $h$.  The loop in question therefore ``shrinks'' downward, as several investigators have observed \citep{Forbes1996,Reeves2008}.  This shrinking occurs relatively slowly, and the loop remains in magnetostatic equilibrium during the process.  We therefore neglect any energetic effect it might have, including the ``collapsing trap'' contributions \citep{Somov1997}.

The energy powering the cooling phase comes primarily from the energy released during the retraction phase.  The majority of this energy (almost $80\%$ in this example) is in perpendicular motion of the flux tube ($K_{\perp}$).  It is not, however, clear how much of this will be dissipated within the flux tube since FMS waves propagate perpendicular to the field.  Without a clear means of resolving this point we restrict consideration to the remaining energy, $0.22\Delta{\cal E}$, in parallel flows and thermal energy.  This flux-tube-confined energy is plotted as a broken line in the lower left panel of \fig\ \ref{fig:erg}.
Perpendicular forces from the underlying field will have little affect on this \citep{Guidoni2011} so it will remain within the flux tube even after it has come to rest.  Eventually we expect all parallel flow to be thermalized, leaving 
$\delta W_T/\delta\Phi=1.8\times 10^{10}\,{\rm erg/Mx}$ in the static tube.  This is the energy powering the cooling phase.

There are few truly simple models of chromospheric evaporation, especially driven by thermal conduction.  Since our main interest lies in the evolution of temperature and density within the loop, we use a zero-dimensional model of the kind originally proposed by \citet{Antiochos1978} and later by \citet{Cargill1995}.  More recently \citet{Klimchuk2008} modified this treatment to account more accurately for the energetic contribution of chromospheric evaporation through the enthalpy flux.  The zero-dimensional loop model, called 
 \EBTEL, yields time-dependent solutions for density and temperature in a loop subjected to additional heating.  
 
Having no spatial resolution, \EBTEL\ cannot capture the kind of retraction-driven energy release we have so far described.  Instead it uses a prescribed, uniform volumetric heating function, $\dot{Q}(t)$, to introduce energy to the loop's plasma.  We use this to represent retraction energy release by specifying a heating profile whose time integral matches $\delta W_T/\delta\Phi$.  Distributed uniformly over a loop $2\,L_{\rm min}=120$ Mm long (excess accounting for the arcade legs) of strength $B=B_g=86$ G, the retraction-produced energy amounts to a density of $117\,{\rm erg/cm}^3$.  This is introduced the \EBTEL\ model with a triangular profile over a 50 second interval. (Changing the assumed duration has little appreciable effect on the solution, which mostly responds to the {\em net} energy input).

\begin{figure}[htbp]
\epsscale{1.0}
\plotone{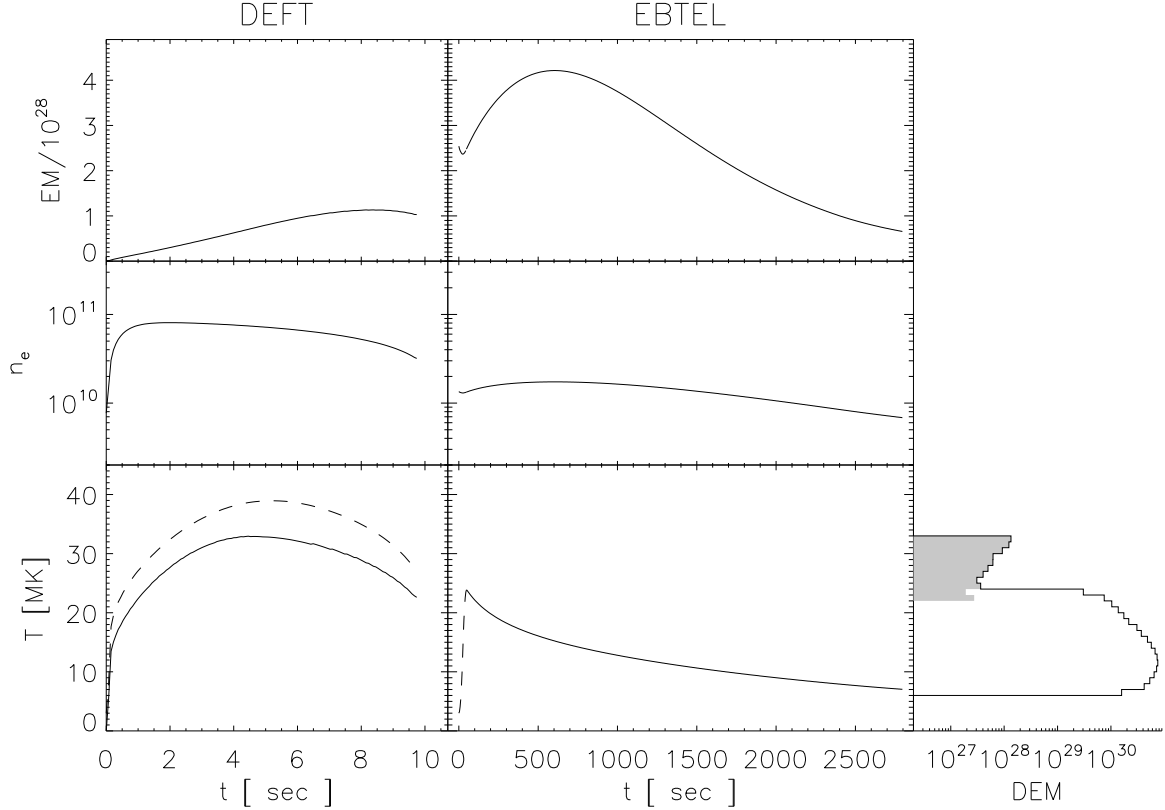}
\caption{Two stages in the evolution of the reconnection episode depicted in \fig\ \ref{fig:erg}.  The left column is the \DEFT\ run covering the retraction through the current sheet.  The other column is the \EBTEL\ solution for evaporation and radiation inside the stationary flux tube.  The temperature (bottom), density (middle) and $EM$ (top, actually 
$\delta EM/\delta\Phi$) are plotted versus time in seconds.  The dashed curve in the lower left panel shows the peak temperature, while the solid shows the $EM$-weighted value.  The graph on the lower right shows the $DEM$ for the combination of both phases.  The grey region shows the $DEM$ of the LTS, from \DEFT.}
	\label{fig:ebtel}
\end{figure}

Figure \ref{fig:ebtel} shows the evolution of both the \DEFT\ solution (left) and the \EBTEL\ solution (right) over their respective simulation times.  While the retraction phase, modeled by  \DEFT, finishes in 10 seconds, the evaporation and cooling phase persists for over half an hour.  The {\em ad hoc} heating phase (dashed curves) drives the \EBTEL\ temperature to 
$T\simeq24$ MK without appreciably affecting the density.  Thermal conduction transports this energy rapidly to the chromosphere where its is partly radiated (the chromosphere is assumed to be a far more efficient radiator than the corona) and the remainder drives evaporative flow, returning the energy through enthalpy flux.  This conductive evaporative phase lasts 600 sec until the coronal density reaches $n_e=2\times10^{10}\,{\rm cm}^{-3}$
and temperature has fallen to $T=15$ MK.  Thereafter the loop cools through coronal radiation and draining flow (downward enthalpy flux).

The two flare phases can be combined into a single differential emission measure ($DEM$) per unit flux change
\be
  DEM_{\dot{\Phi}}(T') ~=~ {1\over \Delta T}\int
  {\delta EM\over \delta\dot{\Phi}}\,\Theta[T(t)-T'-\half\Delta T]\,
  \Theta[T'-\half\Delta T-T(t)]\, dt ~~.
\ee
where the Heaviside functions, $\Theta(x)$, restrict the integral to times the temperature falls within $\half\Delta T$ of the value $T'$.  While it is shown below that this $DEM$ must be interpreted with some care, it is plotted in the lower right of \fig\ \ref{fig:ebtel} against the same temperature axis as the time-profiles.  It has a large peak about $\langle T \rangle = 15$ MK from the cooling phase and a much smaller peak at $\langle T\rangle_{_{SH}}=30$ MK from the \DEFT\ solutions --- the LTS.  The cooling loops (\EBTEL) have a thousand times more $EM$, per unit flux, than the LTS: $7\times10^{31}\,{\rm cm^{-3}\,sec/Mx}$ {\em vs.}
$7\times10^{28}\,{\rm cm^{-3}\,sec/Mx}$.  This is due to their much longer life times, and in spite of their much lower density (\DEFT\ reaches a density 
$n_e=8\times 10^{10}\,{\rm cm}^{-3}$ four times higher than \EBTEL).  Nevertheless, the $DEM$ shows two distinct peaks corresponding to the two phases of post-reconnection evolution.  These, we propose, are the two distinct temperatures detected in flares with SH-LTSs.

To derive observable properties we must combine the single loop simulation into a composite.
If the assumed discrepant flux $\Delta\Psi=3\times10^{21}$ Mx were reconnected over a typical four-minute impulsive phase then the mean flux transfer rate would be 
$\dot{\Phi}\simeq10^{19}\, {\rm Mx/sec}$.  Such a value, equivalent to 100 GV, is characteristic of values found from the rate of flare ribbon motion across magnetograms \citep{Qiu2004,Qiu2005,Longcope2010}.  A single \DEFT\ simulation evolves on far shorter time scales ($\sim 10$ sec, on the left column of \fig\ \ref{fig:ebtel}), so the SH-LTS would consist of a superposition of individual loop-top plugs.  If each reconnection patch transferred $\delta\Phi=10^{19}$ Mx, typical of a post-flare loop in EUV, the impulsive phase would consist of $\sim300$ transfers, occurring at a mean rate of one per second.  Since each loop-top plug remains visible for $\sim10$ sec, roughly $10$ would be visible at any time, to compose the observed LTS.  The emission measure of this composite would be
\be
  \langle EM\rangle_{_{SH}} ~=~\dot{\Phi}\,
  \left({\delta EM\over\delta\dot{\Phi}}\right)_{_{DEFT}} ~\simeq~
  10^{48}\,{\rm cm}^{-3}  ~~.
\ee
characteristic, albeit at the low end, of the range observed in SH-LTSs 
\citep{Lin1981,Nitta1997,Longcope2010,Caspi2010}.  It is evident from the top-left panel of \fig\ \ref{fig:ebtel}, that the $EM$ is very high even at full retraction (i.e.\ the end of the \DEFT\ simultion).  It is therefore possible that $EM$ persists for some time beyond this \citep{Longcope2010,Guidoni2011}.  If this were to happen the observed $EM$ would be larger than our calculation (halted at full retraction) by as much as a factor of two.

The cooling phase occurs over time scales $\sim1000$ sec (right column of \fig\ \ref{fig:ebtel}), much longer than the four-minute flux transfer.  In fact, the emission measure of a single loop peaks at $\delta EM/\delta\Phi_{\rm mx}=4\times10^{28}\,{\rm cm^{-3}/Mx}$, ten minutes after the energy input.  The total emission measure from this phase is not therefore found from \eq\ (\ref{eq:EMphi_dot}) which applies only to rapid evolution.  Instead the net emission measure of the flare as a whole resembles that of the individual curves, peaking at
\be
  {\rm max}\Bigl(\, EM_{_{SXR}}\,\Bigr) ~=~
  \Delta\Psi\,\left({\delta EM\over\delta\Phi}\right)_{\rm mx} ~\simeq ~
  10^{50}\,{\rm cm}^{-3} ~~.
  	\label{eq:GOES_EM}
\ee
GOES would register this cooler plasma, with an $EM$-weighted temperature of 15 MK, as a flux $F_{1\hbox{--}8}\simeq 10^{-4}\,{\rm W/m^2}$ in its low-energy (1--8\AA) channel \citep{Thomas1985}; it would be an X class flare.  

We have thus concluded that, at least for this model flare, the $EM$ of the SXR component is proportional to $\Delta\Psi$, while that of the HXR component is proportional to its time-derivative.  This is basically a re-statement of the Neupert effect \citep{Neupert1968,Dennis1993}.

If the evaporative phase persists far longer than the flux transfer, as we have assumed, 
the $DEM$ observed at any instant will not match the time-integrated curve in the lower right of \fig\ \ref{fig:ebtel}.  The SH-LTS will consist of a super-position of loops in all phases of evolution, so its $DEM$ will resemble the shaded portion of the plot.  The cooling phase, however, evolves much more slowly, so its $DEM$ will be more sharply peaked about the temperature presently found in all the loops; this would differ from the plotted curve, but still result in a distinctly bimodal $DEM$.  The integral under the lower peak would correspond to \eq\ (\ref{eq:GOES_EM}) and will be smaller than the integral under the plotted curve, which matches \eq\ (\ref{eq:EMphi_dot}).  The ratio of areas will be roughly 1:100, as indicated above.  Accounting for the post-retraction phase of the LTS could boost the ratio to 1:50 in this particular example.

The foregoing model X-flare arose directly from a release of $3\times10^{32}\,{\rm erg}$ of magnetic energy through post-reconnection retraction capable of thermalizing  $7\times10^{31}\,{\rm erg}$.  A LTS of $\langle{T}\rangle=30$ MK would be observable above these cooling loops, although its $EM$ would be significantly smaller.  The inferred density of the SH-LTS would be $\sim8\times10^{10}\,{\rm cm}^{-3}$, 
ten times greater than the pre-reconnection density.  

\section{Discussion}

We have herein attempted to show that compression by slow mode shocks, first predicted in the \citet{Petschek1964} model of fast reconnection, is capable of producing high-density, super-hot loop-top emission like that seen in flares.  The density enhancement achieved in these shocks is considerably higher than generally assumed on the basis of steady shock models (Rankine-Hugoniot relations).  This is due to the cooling of the post-shock plasma by thermal conduction during the initial (transient) phase of the post-reconnection retraction.  This effect has been explored using solutions of one-dimensional shock tube equations and two-dimensional thin flux tube equations.  

\begin{figure}[htb]
\epsscale{0.8}
\plotone{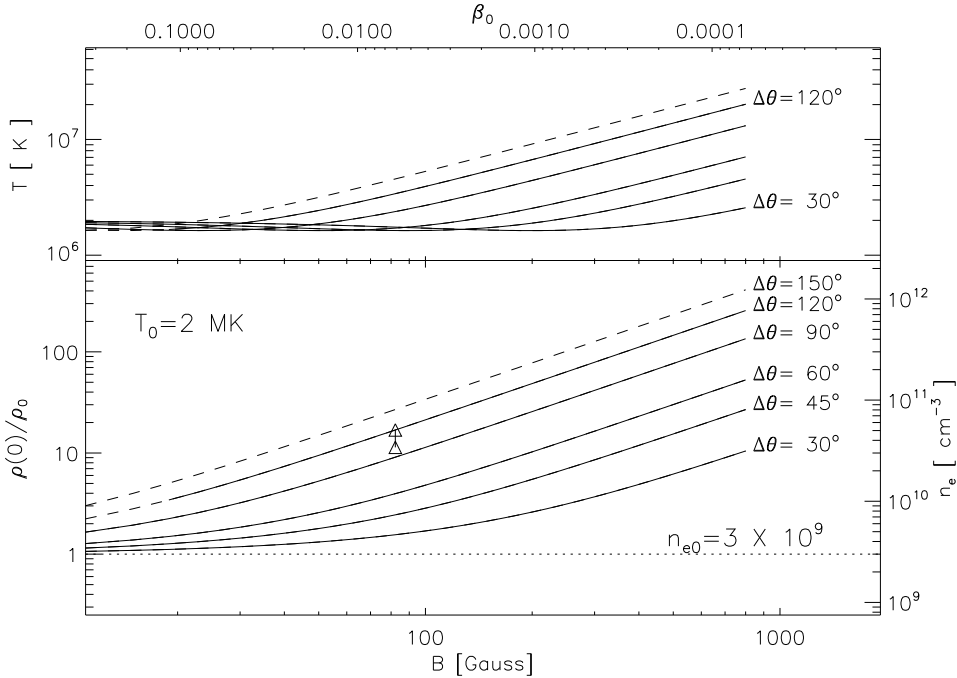}
\caption{Estimates of density enhancement over a range of flare parameters.  For ambient density $n_{e0}=3\times10^{9}\,{\rm cm}^{-3}$ and temperature $T_0=2$ Mk, a range of fields strengths (abscissa) and reconnection angles are used evaluate \eq\ (\ref{eq:rho_ad}), (bottom curve).  The temperature at time of this peak value is plotted in the top panel.  The result from simulation in \fig\ \ref{fig:deft_sweep} ($z_R=h/2$) is shown with a triangle.  The lower triangle is the observed enhancement and the upper triangle (on the curve) is the adiabatic value.}
	\label{fig:bsweep}
\end{figure}

The simplified shock-tube model led us to an estimate of density enhancement, \eq\ (\ref{eq:rho_ad}), in terms of properties of the current sheet and its plasma.  Figure 
\ref{fig:bsweep} uses this relation to show the density and temperature enhancements 
{\em vs.} the strength and angle of the fields separated by the current sheet.  Density enhancements as large as 100 appear plausible for the strongest fields ($B\ga500$ G) and largest angles ($\Delta\theta\ga120^{\circ}$).  This conforms to observational evidence that SH-LTSs occur only when field strengths are extremely high \citep{Caspi2010}.

The shock tube model is a natural abstraction of the thin flux tube model for dynamical evolution following patchy reconnection.  These simplified, time-dependent dynamics turn out to capture the behavior of even classical steady-state MHD reconnection models, provided the reconnecting fields are not perfectly anti-parallel.  In all cases shocks, and corresponding density enhancements, are direct consequences of energy release.  Magnetic energy is released by shortening field lines, which naturally compresses the plasma they link.  Since the field lines shorten at Alfv\'enic speed the plasma is compressed supersonically ($\beta\ll1$) leading to very strong shocks.  

There is quantitative agreement on shock-produced density enhancement between analytic treatments which are steady-state \citep{Soward1982b,Vrsnak2005} and transient 
\citep{Lin1994,Longcope2009}.  This agreement occurs because even in steady-state models, the response of any single field line to an instantaneous topology change is necessarily transient.  All such models have been based on Rankine-Hugoniot jump conditions across the shocks.  We have here shown these steady-shock Rankine-Hugoniot relations are inapplicable to the transient model since a heat front of asymptotic form is never achieved. We fully expect the Rankine-Hugoniot relations will fail in traditional steady-state MHD models as well, since those basically capture the same transient dynamics of each field line.

While analytic treatments might suffer from erroneous assumptions about the shocks, we expect time-dependent MHD solutions to reveal the large density enhancements here predicted.  There have been several two-dimensional simulations of reconnection across a current sheet which included temperature-dependent, field-aligned thermal conduction necessary to our model \citep{Yokoyama1997,Yokoyama1998,Chen1999b,Chen1999}.  While none included a guide field component, the reconnection outflows consisted of heat fronts ahead of isothermal slow shocks --- the same structure predicted by the shock tube model.  The simulations used fairly large initial pressures ($\beta_0\sim0.1$) and thus achieved rather modest shocks:  \citet{Chen1999} observed a fifteen-fold pressure jump, which would be produced by a shock tube inflow at $M_i=2.40$.  That simulation exhibited a five-fold density enhancement across the ``isothermal shock'', twice the maximum achievable by a switch-off sock in steady state but exactly matching shock tube prediction \eq\ (\ref{eq:rho_ad}).  Since the anti-parallel reconnection
($\Delta\theta=180^{\circ}$) simulated by \citet{Chen1999} falls outside the validity of the shock-tube model, this agreement must be regarded as fortuitous.

A more direct treatment of the full problem, including a guide-field, using the full set of MHD equations could bypass the limitations of the thin flux tube model.  This model requires that the plasma-$\beta$ be small both before and after reconnection.  For negligible initial values ($\beta_0\ll1$), the post shock value approaches the limit, $\beta_2\simeq (16/3)\sin^4(\Delta\theta/4)$, 
\citep{Longcope2009} which exceeds $0.75$ for $\Delta\theta>150^{\circ}$.  Thus reconnection at more acute angles, where $M_i$ could be large, cannot be treated using our thin flux tube model.  Indeed, there is observational evidence \citep{Caspi2010} that many SH-LTSs achieve a plasma pressure limited by the magnetic pressure ($\beta_2\simeq1$).  We therefore hope in the future to use full MHD simulations to explore this reconnection regime.

In both steady or transient scenarios, the total emission measure of the hot reconnection outflow will be proportional to the mean reconnection rate $\dot{\Phi}$ according to 
\eq\ (\ref{eq:EMphi_dot}).  In traditional steady-state models this is equivalent to an electric field along the X-line \citep{Petschek1964,Vasyliunas1975,Soward1982}, directly related to reconnection microphysics.  This contrasts with patchy reconnection scenarios \citep{Klimchuk1996,McKenzie1999} where the local electric field determines the time required to reconnect each patch, but $\dot{\Phi}$ is proportional to the rate patches are produced, an unrelated quantity.  While it is important to resolve this ambiguity if we are to fully understand solar flares, both scenarios lead to the same relation between $\dot{\Phi}$ and the flare's $EM$ for physical reasons outlined above. Moreover, the mean reconnection rate $\dot{\Phi}$ can be directly measured using observation of flare ribbon motion \citep{Forbes1984,Fletcher2001,Qiu2002}.  
The values typically obtained are consistent with the 
$EM$ of typical SH-LTSs.  Such measurements have been interpreted as X-line electric fields by applying the two-dimensional steady state model \citep{Poletto1986}.  Doing so neglects the complexities of flare ribbons and magnetic fields, which seem to suggest a less ordered, more patchy reconnection process \citep{Fletcher2004,Longcope2007,Qiu2009}.

The constant of proportionality, $\delta EM/\delta\dot{\Phi}$, can be derived from global, observable quantities.  It consists of a factor, given in \eq\ (\ref{eq:EM_fctr}), depending on pre-flare conditions and the vertical extent of the pre-flare current sheet, multiplied by a dimensionless factor depending primarily on the angle 
$\Delta\theta$  between reconnecting field lines which determines the inflow Mach number $M_i$ according to \eq\ (\ref{eq:Mi}). 
This simple dependence can be illustrated using the 
$40$ MK SH-LTS studied by, \citet{Longcope2010}, which had 
a constant $\delta EM/\delta\dot{\Phi}\simeq3\times 10^{29}$: it had an emission measure  
$EM\simeq3\times 10^{48}\,{\rm cm}^{-3}$ during reconnection at a mean rate
$\dot{\Phi}\simeq10^{19}$ Mx/sec, measured using ribbon motion.  
Its thirteen-fold temperature increase, from $T=3$ MK to $T=40$ MK, requires a shock of $M_i\simeq4.7$, as would occur for reconnection between fields differing by 
$\Delta\theta\simeq90^{\circ}$ \citep[roughly consistent with the pre-flare magnetic 
model of][]{Longcope2010}. Reading from \fig\ \ref{fig:smry}, 
this should produce an $EM\sim0.3$ times the factor in \eq\ (\ref{eq:EM_fctr}), which must therefore be $10^{30}\,{\rm sec\,Mx^{-1}\,cm^{-3}}$.  A 
current sheet $h=30$ Mm high with guide field $B_g=200$ G \citep[similar to those of][]{Longcope2010} would require a pre-reconnection density (density inside the current sheet on the un-reconnected field lines) $n_{e0}=2\times10^{10}\,{\rm cm}^{-3}$, to produce the observed LTS.
This is lower than the value quoted by \citet{Longcope2010} due primarily to density enhancement above the Rankine-Hugoniot value.   According to \fig\ 
\ref{fig:smry} the loop-top density in the SH source would be 
$n_e\simeq3\times10^{11}\,{\rm cm}^{-3}$, five times greater than the maximum 
permitted by Rankine-Hugoniot relations.

Perhaps the greatest puzzle raised by the collisional model explored here is how it might relate to flares (or flare phases) with significant non-thermal populations.  Super-hot sources often appear after the tell-tale power-law HXR spectrum has vanished \citep{Alexander1997,Jiang2006,Longcope2010}, so the collisional and collisionless processes may occur separately.  It still seems reasonable to assume the same underlying mechanism is responsible for energy transfer in both cases.  Here we have shown how reconnection can thermalize a significant fraction of stored magnetic energy through slow magnetosonic shocks, assuming sufficiently high collision rates.  Could a related process, at lower densities, produce the non-thermal populations observed at other times and in other flares?  Except in cases of anti-parallel reconnection the slow shock is essentially a {\em parallel shock}, whose collisionless manifestation is still poorly understood 
\citep[see][for overviews of the topic and different possible approaches]{Parker1961,Sagdeev1966,Stone1985,Quest1988,Khabibrakhmanov1993}.  
This seems to be an avenue worth pursuing given that the present model outlines a complete energetic chain from current sheet to post-flare loops.

In any event, the large densities within the SH-LTS are sufficient to thermalize the electron population, thereby explaining the absence of a non-thermal population when they are present.  A troubling discrepancy occurs in trying to explain how bulk kinetic energy, produced at the rotational discontinuities and carried almost entirely by ions, can be transformed to electron thermal energy.  Even at the high densities observed the classical rate of ion-electron collision is too low for this transfer to be effective \citep{Longcope2010b}.  Since all measurements available are of {\em electron} temperatures in flares, they must be heated somehow, in spite of this theoretical hurdle.  This would not be the first instance where actual collision rates greatly exceeded those of classical Coulomb interactions.  Indeed, the resolution of this paradox may also provide a clue to the production of non-thermal electrons in cases of lower density.

\acknowledgements

This work was supported  by NSF and DOE under a joint grant.


\end{document}